\documentclass{article}
\usepackage[utf8]{inputenc}
\usepackage{graphicx}
\usepackage{authblk}
\usepackage[backend=biber, style=ieee]{biblatex}
\usepackage[a4paper, left=2cm, right=2cm, top=2cm, bottom=2cm]{geometry}
\usepackage[colorlinks=true, linkcolor=blue, citecolor=blue]{hyperref}
\usepackage{caption}
\usepackage{subcaption}

\title{A Methodology for CHF Prediction in VVER Rod Bundles}
\author[1]{Mohamed AbdulHameed}
\author[1]{Aly Shaaban}
\author[1]{Hussein Gamal}
\author[1*]{Ayah Elshahat}
\affil[1]{Department of Nuclear and Radiation Engineering, Faculty of Engineering, Alexandria University, Egypt}
\affil[*]{Corresponding author}
\date{}

\begin{document}

\maketitle

\begin{abstract}

The critical heat flux (CHF) is an important thermal-hydraulic parameter that must be determined during the design of water-cooled reactors to ensure safety of operation. In this paper, a methodology is proposed to predict the CHF in VVER\footnote{VVER is a transliterated acronym for ``vodo-vodyanoi energetichesky reaktor'' which literally means ``water-water power reactor'' \cite{VVER}.} rod bundles. A whole-reactor model of VVER-1000 is implemented in RELAP-SCDAPSIM, and, based on the bundle-average approach, the parameters of the hot channel are used to assess four CHF predictors (i.e., the 2006 Groeneveld look-up table (LUT), and the Bowring, W-3, Biasi, and OKB-Gidropress correlations) upon comparison with the 2011 Bobkov LUT. A brief critique is given of the correction factors for both the Groeneveld and Bobkov LUTs. The effects of channel diameter and non-uniform axial heat flux on the predicted values of CHF in subcooled conditions are also studied. The performance and time complexity of the standard and Bourke trilinear interpolation algorithms are assessed for use with LUTs in safety codes. It has been concluded that the non-uniform axial heat flux does not affect the CHF in subcooled conditions and that the Groeneveld diameter factor exponent gives better predictability of the CHF in VVER rod bundles over the Tanase and Wong exponent correlations. Additionally, the Groeneveld LUT has been found to nearly reproduces the Bobkov LUT CHF values when used in conjunction with the Bobkov heated length factor. Values of MDNBR equal to 1.83 and 1.96 can be used as thermal limits at 12\% overpower when the Bowring and W-3 correlations, respectively, are used in subcooled conditions. The standard and Bourke algorithms nearly exhibit the same average CPU time. The Bourke algorithm, however, is less affected by noise. Further research is needed to ensure the smoothness of the Bobkov LUT and to derive reactor type-dependent correction factor formulas for use with the Groeneveld LUT.

\end{abstract}

\section{Introduction}
\label{intro}

An important thermal-hydraulic (TH) parameter that must be determined during the design of water-cooled reactors is the critical heat flux (CHF). Four approaches exist for predicting the CHF in rod bundles:
\begin{enumerate}
    \itemsep -0.3 em
    \item Analysis of TH conditions of each coolant subchannel in the rod bundle (i.e., the subchannel approach \cite{Kao1983}). 
    \item Analysis of average TH conditions over the whole bundle (i.e., the bundle-average approach \cite{Kao1983}).
    
    \item The enthalpy imbalance approach \cite{IAEA2001}, which has been proposed as an alternative to the subchannel analysis. The enthalpy imbalance approach defines the difference in enthalpy rise rate (Eq. \ref{Eq:dH}) among a bundle's subchannels as a function of \textit{(a)} fuel rod's gap to diameter ratio, $(s-d)/d = \delta / d$, where $s$ is the fuel rod pitch, and $d$ is the fuel rod diameter, and \textit{(b)} a quality imbalance, $\Delta x$, for that bundle. $\Delta x$ is defined as the difference in qualities between the average bundle quality and the maximum bundle subchannel quality for a given cross-section. A general expression for $\Delta h$ is yet not available; only empirical expressions proposed for specific bundle geometries \cite{IAEA2001}.
    \item The ``steaming'' approach proposed by Kachur \cite{Kachur2019} that applies methods of random processes to predict, at early stages, the occurrence of abnormal heat exchange in a reactor core based on a small number of parameters. This approach, originally developed for VVER monitoring and control, should minimize indirect calculations and over-reliance on empirical formulas. Due to its recency, Kachur's approach hasn't yet been used to predict CHF in rod bundles.
\end{enumerate}

\begin{equation}
\Delta h = f \left( \delta / d, \ \Delta x_{\mathrm{max}} \right)
\label{Eq:dH}
\end{equation}

CHF prediction methods are based on experiments in either round tubes or rod bundles, and are presented in one of two forms: \textit{(a)} look-up tables (LUTs), or \textit{(b)} empirical correlations. Compared to empirical correlations, LUTs possess higher accuracy and a wider range of validity \cite{IAEA2001, Cheng2003}. Regardless of their form, CHF ``predictors'' for low and subcooled conditions fall into one of three types \cite{Hejzlar1996}:
\begin{enumerate}
    \itemsep -0.3 em
     
    \item Predictors that depend completely on local conditions (``local'' predictors). These predictors have the form of Eq. \ref{Eq:DSM}, or Eq. \ref{Eq:DSM2} if the heated length is taken into account. Examples of this type include the 2006 Groeneveld LUT \cite{Groeneveld2007}, the 2011 Bobkov LUT \cite{Bobkov2011}, and the Biasi correlation \cite{Kumamaru1989}.
    
    \item Predictors that incorporate a heat balance of the form of Eq. \ref{Eq:HB} (``holistic'' predictors), where $P_h$ is the heated perimeter, $G$ is the mass flux, $S$ is the channel's cross-section area, $h_{\mathrm{fg}}$ is the latent heat, and $T$ is the heat-balance axial heat flux parameter defined by Eq. \ref{Eq:TB} \cite{Hejzlar1996}. These predictors have the form of Eq. \ref{Eq:HBM}. Examples of this type include the Bowring correlation \cite{Kao1983}.
    
    \item ``Semilocal'' predictors. These are hybrid predictors that take into account inlet conditions (e.g., $x_{\mathrm{in}}$ or $h_{\mathrm{in}}$) but do not incorporate a full heat balance as in holistic predictors. Some of them also take into account heat length effects by introducing the relative heated length, $z/D_h$. A typical example of this type is the W-3 correlation \cite{Todreas2012}.
\end{enumerate}

\begin{equation}
\mathrm{CHF} = f \left( G, \ x, \ p, \ D_h \right)
\label{Eq:DSM}
\end{equation}

\begin{equation}
\mathrm{CHF} = f \left( G, \ x, \ p, \ D_h, \ L \right)
\label{Eq:DSM2}
\end{equation}

\begin{equation}
q(z) = \frac{x-x_{in}}{T(z) P_h z} G S h_{\mathrm{fg}}
\label{Eq:HB}
\end{equation}

\begin{equation}
T(z) = \frac{1}{q(z) z} \int_{0}^{z} q(z') dz'
\label{Eq:TB}
\end{equation}

\begin{equation}
\mathrm{CHF} = f \left[ G, \ x_{\mathrm{in}} \left( \mathrm{or} \ h_{\mathrm{in}} \right), \ p, \ D_h, \ z \left( \mathrm{or} \ \frac{x-x_{\mathrm{in}}}{q} \right) \right]
\label{Eq:HBM}
\end{equation}

Two thermal limits exist to prevent the boiling crisis in water-cooled reactors. For PWRs, the ratio of CHF to actual wall heat flux, i.e., the DNB ratio (DNBR), will be kept below 1.3 at 112\% of reactor nominal power \cite{Todreas2012}. For BWRs, the ratio of critical power to average power, i.e., the critical power ratio (CPR), will be kept below 1.2. CPR takes into account the effect of upstream conditions, e.g., axial flux non-uniformity, on the downstream CHF, whereas DNBR depends completely on local conditions \cite{Todreas2012}. 

\begin{figure}[h]
    \centering
    \includegraphics[width=0.6\textwidth]{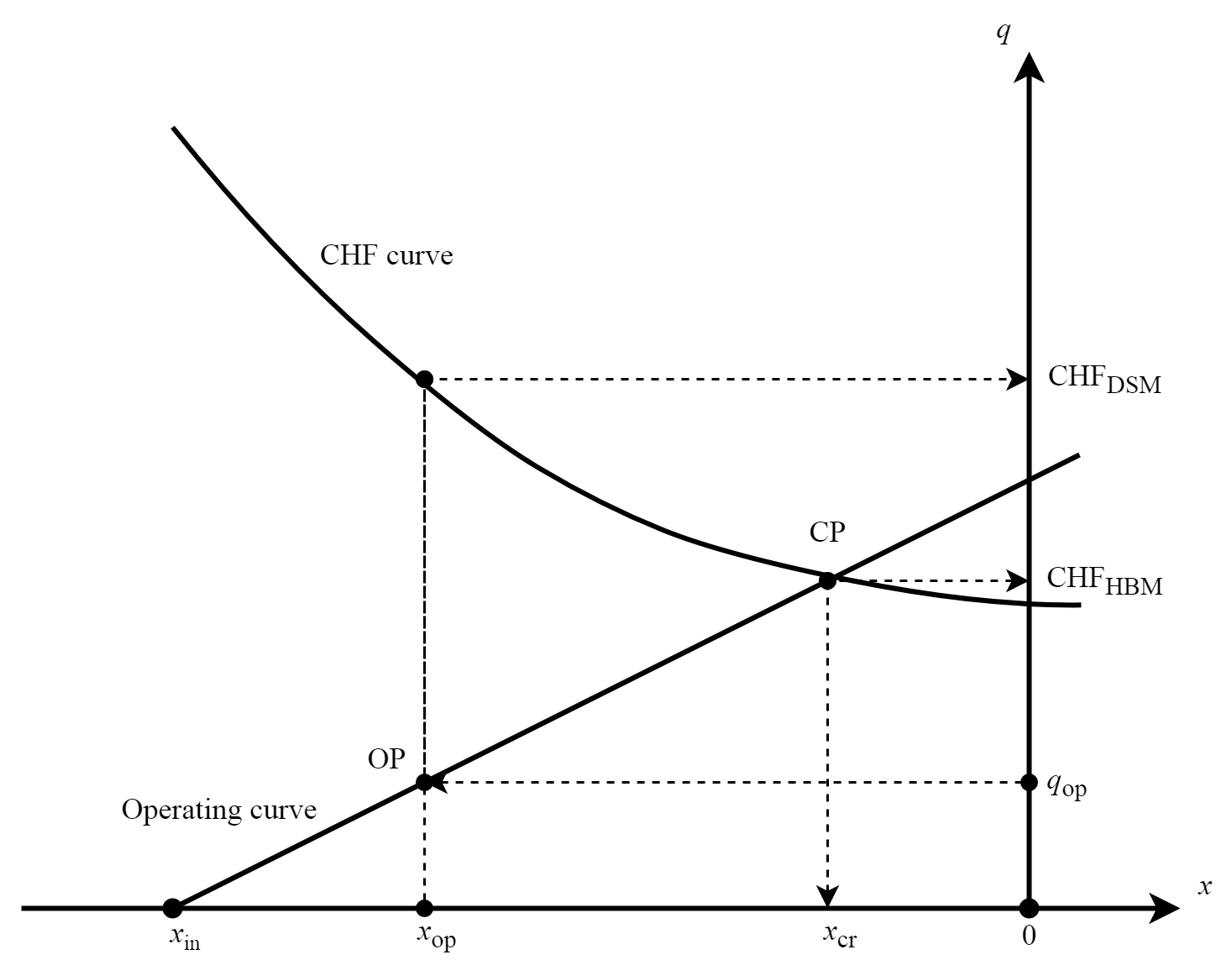}
    \caption{Methods for obtaining the CHF illustrated in $q$-$x$ space at a fixed loaction, $z$. The operating curve is based on the heat balance (Eq. \ref{Eq:HB}). The CHF curve may represent any CHF predictor of any type. CP stands for critical point. OP stands for operating point. (Reproduced from Ref. \cite{Hejzlar1996})}
    \label{Fig:HBMvsDSM}
\end{figure}

As illustrated in Figure \ref{Fig:HBMvsDSM}, there exist two methods to obtain the CHF, and consequently the DNBR \cite{Hejzlar1996}:

\begin{enumerate}
	\itemsep -0.3 em
	
	\item The direct substitution method (DSM). In this method, the operating heat flux, $q_{\mathrm{op}}$, is substituted into the heat balance (Eq. \ref{Eq:HB}) to calculate the operating quality, $x_{\mathrm{op}}$. $x_{\mathrm{op}}$ is then substituted into the CHF correlation to obtain the CHF at the location of interest. Safety codes (e.g., RELAP-SCDAPSIM, the tool of our analysis) are inherently designed to use the DSM \cite{Hejzlar1996}.

	\item The heat balance method (HBM). The idea of the HBM is to gradually increase the channel's power until a boiling crisis occurs at the location of interest, that is, until the operating curve, $q(x)$, intersects with the CHF curve, $\mathrm{CHF}(x)$, in the $q$-$x$ space. Experimentally, this is achieved by computer-controlled electric heaters and a thermocouple that detects a sudden increase in temperature indicating boiling crisis occurrence \cite{Yang2006}.
\end{enumerate}
	
For local and semilocal predictors, the HBM can be computationally implemented by two procedures:
	\begin{enumerate}
	\itemsep -0.3 em
	\item Solving Eq. \ref{Eq:op-CHF} for $x_{\mathrm{cr}}$, which requires an iterative solution. This procedure, however, is only possible for isolated subchannels for which energy balance is valid.
	\item Incrementally increasing channel power until critical conditions are achieved at the location of interest. This procedure is possible for both isolated subchannels and rod bundles. For rod bundles, however, it must be performed by a subchannel code that handles the crossflow between different subchannels \cite{Hejzlar1996}. It is important to note that this is the same procedure used to calculate the critical power ratio (CPR) \cite{Hejzlar1996}. That is, in this case, the DNBR and the CPR are equivalent.  
	\end{enumerate}

For holistic predictors, no iterative solution is necessary because Eq. \ref{Eq:op-CHF} is just the heat balance, which is inherent to correlations of this type.

\begin{equation}
\frac{x-x_{\mathrm{in}}}{T(z) P_h z} G S h_{\mathrm{fg}} - f \left( G, \ x, \ p, \ D_h \right) = 0
\label{Eq:op-CHF}  
\end{equation}


    
Computationally, procedures for either the DSM or the HBM have to be repeated at each ``spatial increment'' along the simulated channel. Hejzlar and Todreas \cite{Hejzlar1996} have demonstrated that while local and semilocal predictors give higher DNBR values for the DSM than for the HBM, holistic predictors give the same DNBR value when used with both methods. For semilocal predictors, however, the difference between the two DNBR values is much smaller \cite{Hejzlar1996} because predictors of this type take into account inlet conditions. Hejzlar and Todreas have concluded that the more the weight given to inlet conditions, the less the difference \cite{Hejzlar1996}. LUTs, considered the ``standard'' CHF predictor, are implemented in modern safety codes using the DSM \cite{RELAPModels, Hejzlar1996}. To overcome their ``locality'' and account for channel geometrical anomalies, LUTs are made ``semilocal'' by introducing empirical correction factors. The accuracy of this approach depends on the accuracy of the correction factor formulas.


The goals of this paper are to:
\begin{itemize}
    \itemsep -0.3 em
    \item Propose a consistent methodology to predict the CHF in VVER rod bundles,
    \item Assess the performance of the three CHF predictor types using the DSM,
    \item Enhance the predictability of the Groeneveld LUT of the CHF in VVER rod bundles,
    \item Study the effect of non-uniform axial heat flux (cf. Section \ref{CHFTrend}) on the CHF in subcooled conditions,
    \item Determine which diameter factor exponent (cf. Section \ref{GRO}) should be used with the Groeneveld LUT when predicting the CHF for VVER rod bundles, and
    \item Compare two trilinear interpolation algorithms (cf. Section \ref{INTERPOL}) for use with LUTs. 
\end{itemize}

RELAP-SCDAPSIM MOD 4.0, developed by Innovative Systems Software (ISS), is used to model VVER-1000. The input deck for VVER-1000 was provided to the Nuclear and Radiation Engineering Department of Alexandria University by ISS \cite{InnovativeSystemsSoftware} and was further expanded and modified by the authors. The ISS input deck is based on the input deck developed by the Institute for Nuclear Research and Nuclear Energy, Bulgarian Academy of Sciences (INRNE-BAS) \cite{Vryashkova2015} whose reference plant is the Kozloduy Nuclear Power Plant (KNPP) Unit 6 (V-320 model). The INRNE-BAS input deck has been validated against many benchmarks (most of these analyses are proprietary) \cite{Allison2010}, the most recent of which is the simulation and integrated uncertainty analysis of the OECD/NEA VVER-1000 coolant transient benchmark (V1000CT-2) \cite{Spasov2021}.

A steady-state simulation of VVER-1000 at 112\% of nominal power is performed by RELAP-SCDAPSIM. The CHF is then calculated for the hot channel using the Groeneveld and Bobkov LUTs, and the W-3, Biasi, Bowring, and OKB-Gidropress correlations. For reasons detailed in Section \ref{Bobkov}, the Bobkov LUT is treated as the ``benchmark'' against which other predictors are assessed. Being the most recent LUT, a comprehensive comparison among the 2011 version of the Bobkov LUT and other CHF predictors hasn't been performed, as far as the authors are aware.

\section{RELAP-SCDAPSIM model description}

RELAP-SCDAPSIM comprises two codes: RELAP5, and SCDAP. RELAP5 models hydrodynamic volumes, piping heat structures, control systems, and neutron kinetics, whereas SCDAP models core heat structures. RELAP predicts system thermal-hydraulics, control system interaction, etc., while SCDAP offers better predictability of core thermal phenomena during accidents \cite{Sanchez-Espinoza2009} as well as steady states.

\subsection{Pressure vessel}
\label{Sec:PV}

\begin{figure}[h]
    \centering
    \includegraphics[width=0.75\textwidth]{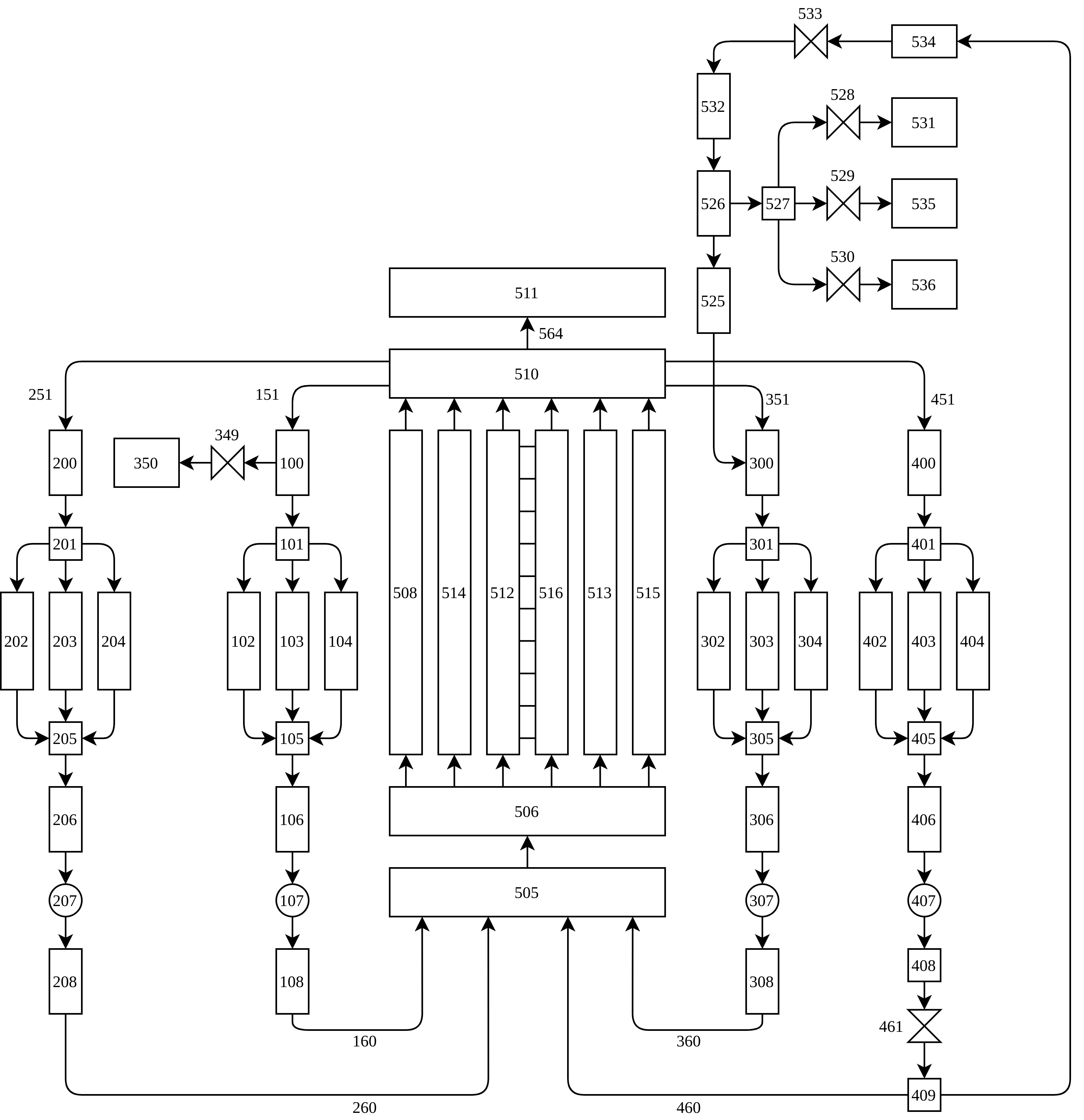}
    \caption{VVER-1000 pressure vessel and primary circuit component schematic (Circles represent pumps.)}
    \label{Fig:NP}
\end{figure}

The component schematic of the VVER-1000 pressure vessel and the primary circuit is shown in Figure \ref{Fig:NP}. The core is modeled with 5 channels (components 512, 513, 514, 515, and 516) and ten axial nodes per channel. The flow area of each fuel channel can be found in Table \ref{Tab:SCDAP}. Channel 516 models the hottest channel, and channel 512 models the second hottest channel. Component 508 is a ten-node channel that models bypass flow.

Asymmetric thermal-hydraulic conditions can occur in the downcomer during Emergency Core Cooling (ECC) injection, which can be predicted by RELAP5 if a multi-component downcomer is modeled, i.e., a component connected to each loop, and the four components are connected by cross-flow junctions \cite{Sanchez-Espinoza2009, Hendrix1992, Grudev2004}. For our purposes, however, the downcomer (component 505) is modeled using a 4-node annulus, in which each loop is connected to a node. Components 506, 510, and 511 model the lower plenum, upper plenum, and upper head, respectively. Component 564 describes a leakage path between the core and the upper head through the control elements.

For the model to be accurate enough for CHF prediction, three aspects need to be ensured:
\begin{enumerate}
    \itemsep -0.3 em
    \item Modeling of in-core fluid mixing due to turbulence and geometric peculiarities of rod bundles. This can be accomplished using cross-flow junctions. To fulfill this purpose, axial nodes of channels 516 and 512 were connected by cross-flow junctions. Input data for the cross-flow junctions were based on the recommendations of Ref. \cite{KoreaInstituteofNuclearSafetyKINS2000}. 
    \item Modeling core inlet losses. For this purpose, a loss coefficient\footnote{The loss coefficient of a component is defined as the ratio of dissipated and kinetic energies in this component \cite{Herwig2010}.} of 0.005 has been assigned to components 505 (the downcomer) and 506 (the lower plenum). In addition, the wall friction model, as well as the interphase friction model, have been activated for the lower plenum. Details about these models can be found in Ref. \cite{RELAPModels}.
    \item Verifying model fidelity in predicting in-core TH parameters, because the CHF is primarily a function of those parameters. This is done in Section \ref{Sec:Val}.
\end{enumerate}

\subsection{Primary circuit}

Technically speaking, the pressure vessel is part of the primary circuit. However, it incorporates such many details that we preferred to treat it separately. The rest of the primary circuit is described in this section.

The primary circuit consisting of the piping system (loop 1: components 151, 100, 101, 105, 106, and 108), the pumps (components 107, 207, 307, and 407), and the steam generator (SG) tubes (SG 1: volumes 102, 103, and 104). To model heat exchange, RELAP5 heat structures are defined for both the primary and secondary sides of the SG. The Westinghouse pump model, built in RELAP5, is used to model the pumps. In addition, the pressurizer (PRZ), i.e., component 526, with the four groups of heaters is included in the model. The PRZ’s spray line (component 532) is connected to loop 4, whereas its surge line (component 525) is connected to loop 3. A set of relief valves are also connected to the pressurizer. The heat input of the pressurizer heaters is based on data reported in the ISS input deck \cite{InnovativeSystemsSoftware} and is modeled using 4 trip-controlled general tables. The spray valve opens when the PRZ pressure is greater than 15.7 MPa; heaters activate sequentially at PRZ pressures of 15.5 MPa, 15.3 MPa, 15.1 MPa, and 14.9 MPa; and relief valves open when the pressure in the pressurizer is greater than 15.72 MPa. These pressure setpoints were not included in the original ISS input deck and have been chosen conveniently by the authors due to the lack of precise input data.


To include a simplified sensitivity analysis in the validation process, the number of nodes for some components in loop 4 has been changed to examine to what extent the calculated TH parameters are sensitive to changes in the component schematic. For instance, the cold leg in loop 4 comprises only 2 nodes, whereas the cold legs of the other three loops comprise 4 nodes. It is reasonable to assume that the sensitivity of out-of-core components extends to in-core components. A more thorough analysis, however, would select a range for the number of nodes in core channels and apply the methodology for each number in this range, which would require a separate research paper.  

\subsection{Secondary circuit}

\begin{figure}[h]
    \centering
    \includegraphics[width=0.75\textwidth]{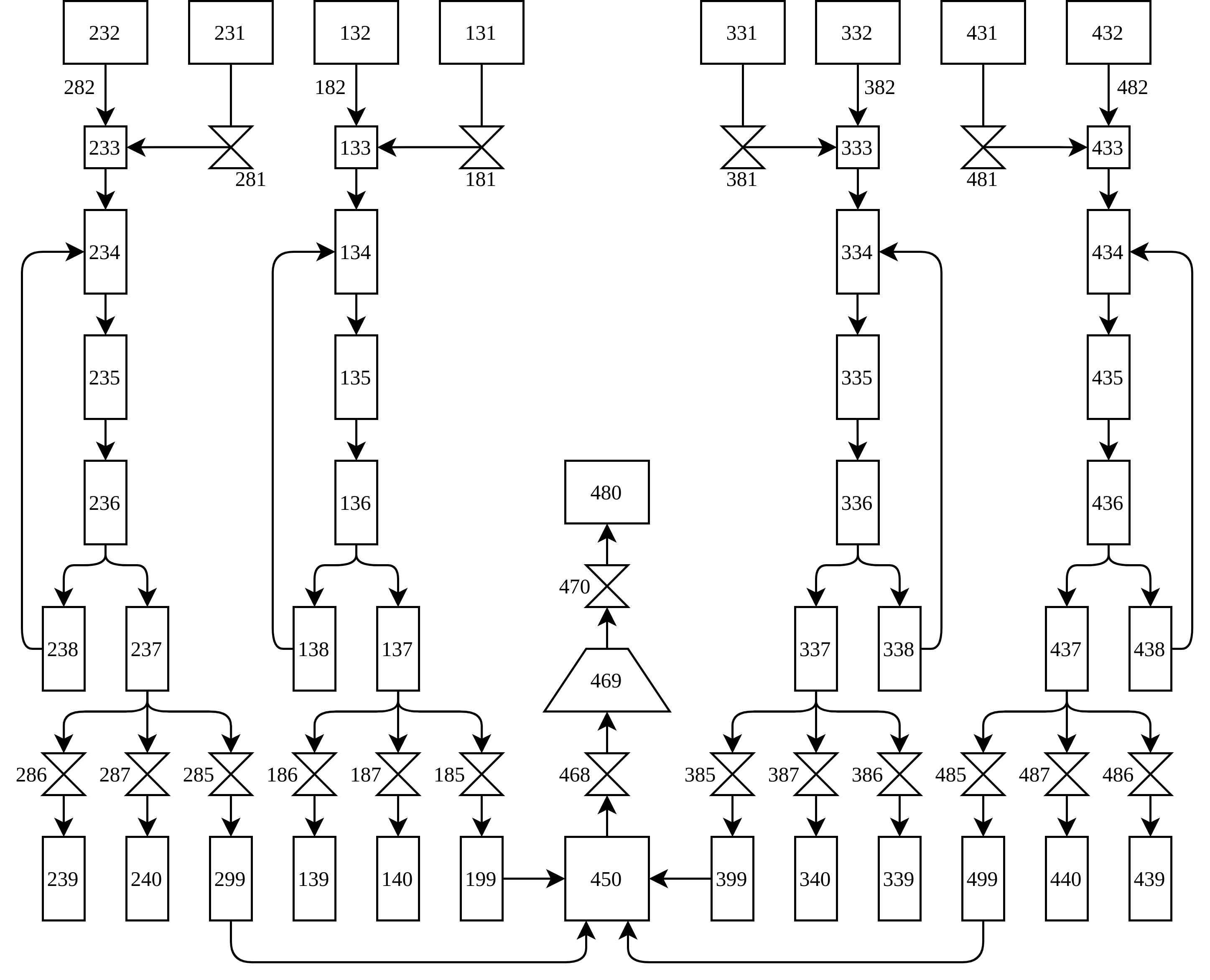}
    \caption{VVER-1000 secondary circuit component schematic}
    \label{Fig:NS}
\end{figure}

The component schematic of the VVER-1000 secondary circuit is shown in Figure \ref{Fig:NS}. The feedwater system is modeled using time-dependent volumes (components: 131, 231, 331, and 431) providing a constant mass flow with a predefined coolant temperature through valves 181, 281, 381, and 481, and is treated as a boundary condition. The auxiliary feedwater (AFW) system is also modeled with time-dependent volumes (components: 132, 232, 332, and 432) connected to the SG's secondary side inlet through time-dependent junctions 182, 282, 382, and 482.

The secondary side of the SGs (SG 1: components 133, 134, 135, 136, and 138) and the steam lines are modeled with minimum sufficient detail. The steam line includes the common header (component 450), the turbine stop valve (component 468), and the steam dump to atmosphere (BRU-A) valves. Steam dump to condenser (BRU-K) valves are not modeled because the condenser is modeled by the time-dependent volume 480, and treated as a boundary condition.

\subsection{Neutron kinetics}

The model uses point kinetics to model the time-dependent behavior of neutrons. Feedbacks due to moderator density, void fraction weighted moderator temperature, and fuel temperature are assumed to be independent. The concentrations of fission products and actinides are assumed to have reached saturation. The reactor is assumed to operate at a nominal power of 3000 MW with zero initial reactivity.

\subsection{SCDAP input}

UO$_{2}$ fuel rods are modeled with average burnup of 21 MWd/kg, and fuel-to-clad gap of 83-$\mu$m width \cite{InnovativeSystemsSoftware}. The gap in each fuel rod is modeled to contain 9.15 $\times 10^{-5}$ kg of helium at a pressure to be calculated during the iteration procedure \cite{InnovativeSystemsSoftware, Hohorst2012}. Zircaloy, the default PWR clad material, has been used in the SCDAP input instead of alloy E-110 (i.e., Zr-1\% Nb), the default VVER clad material \cite{IAEA2011, Todreas2012, Hohorst2012}. The reason for this is that Zr-1\% Nb is not defined within the SCDAP code \cite{Hohorst2012}. This approximation is sufficient for our purposes because the thermal properties of both Zircaloy and Zr-1\% Nb do not differ significantly \cite{Todreas2012}.

As stated previously, the reactor core is described with five channels and ten axial nodes per channel. SCDAP requires that an axial power fraction (APF) be input for each axial node. Axial power variation was assumed to be identical in the five channels. Table \ref{Tab:APF} shows the APFs for each node. Although axial power variation must be included, radial power variation can be neglected for fuel rods \cite{Duderstadt1976}. Consequently, the radial power fraction was assumed to be unity all over the core. Each channel contains an SCDAP fuel rod heat structure and an SCDAP control rod heat structure. Table \ref{Tab:SCDAP} lists the numbers of fuel rods and control rods included in each channel. Fuel rod heat structures are modeled using 6 radial nodes. However, control rod heat structures are modeled using only 2 nodes due to an SCDAP limitation. The SCDAP code also requires an input of a radial temperature distribution at each axial node. A uniform initial temperature distribution was used for both fuel rods and control rods as these temperatures will adjust when a steady-state calculation is performed. VVER-1000 fuel assemblies comprise 14 spacer grids \cite{IAEA2003}. Due to computing power limitations, only 7 spacer grids were modeled by SCDAP. SCDAP model of the spacer grids is only concerned by the spacer gird-clad chemical interaction during severe accidents \cite{TheSCDAP-RELAP5DevelopmentTeam1997} and doesn't affect steady-state CHF calculations. It is only included for completeness.

\begin{table}[h!]
    \centering
    \begin{tabular}{c|cccccccccc}
    \hline
    Node & 1 & 2 & 3 & 4 & 5 & 6 & 7 & 8 & 9 & 10 \\
    \hline
    APF & 0.589 & 1.129 & 1.154 & 1.114 & 1.089 & 1.079 & 1.079 & 1.074 & 1.014 & 0.679 \\
    \hline
    \end{tabular}
    \caption{Core channels' axial power factors \cite{InnovativeSystemsSoftware}}
    \label{Tab:APF}
\end{table}

\begin{table}[h!]
    \centering
    \footnotesize{
    \begin{tabular}{c|c|c|c}
    \hline
    Channel & Fuel rods & Control rods & Modeled flow area [m$^2$] \\
    \hline
    512     & 5909      & 234 & 0.468084 \\
    513     & 13062     & 540 & 1.034712 \\
    514     & 9330      & 540 & 0.739080 \\
    515     & 11196     & 540 & 0.866896 \\
    516     & 11196     & 576 & 0.866896 \\
    \hline
    \end{tabular}
    }
    \caption{Fuel rods, control rods, and flow area of each fuel channel \cite{InnovativeSystemsSoftware}}
    \label{Tab:SCDAP}
\end{table}

\section{Model validation}
\label{Sec:Val}

Geometric validation of the input deck has been performed independently by comparing the component schematic against VVER-1000's actual components and systems as documented by ROSATOM and the IAEA \cite{ROSATOM, IAEA2011}. It has been found that the primary system model shows a satisfactory degree of accuracy. On the other hand, the secondary system is only approximate. It's a non-complete loop that is based on pre-defined boundary conditions that duplicate the reactor's steady-state. Consequently, while the steady-state parameter values of the secondary circuit might be reliable, its transient behavior will be treated with care. No transient calculation is performed in this paper, however, and the input deck is satisfactory for our purposes.

Input deck parameter values have been validated by performing a steady-state simulation for 2000 seconds and comparing the results with measured VVER-1000 plant data found in Ref. \cite{Grudev2004}, which contain measured values at Unit 6 of the Kozloduy nuclear power plant (KNPP), Bulgaria, which is of the V-320 model \cite{ROSATOM}. The results of input validation are presented in Tables \ref{Tab:PerChannel}$-$\ref{tab:PerSLoop}. All parameter values have been approximated to 4 significant figures.

By 1250 seconds, all calculated parameters have reached a constant value. That is, the steady state is achieved by 1250 seconds. The last parameters to converge in our analysis were the primary circuit pressures. This is likely due to the approximations used in defining the pressurizer control logic. In the analysis by Grudev and Pavlova, the steady state has been achieved in only 60 seconds \cite{Grudev2004}. This much shorter time is likely due to Grudev and Pavlova's reliance on measured plant parameters as initial conditions in their input deck, whereas, on the other hand, the ISS input deck frequently uses 0 and 1 as initial conditions when input data are not available, which takes RELAP-SCDAPSIM longer time to converge to a steady-state value. 

When computing relative errors for per-channel values, average values over all the channels were used. This is based on the fact that channels are only computational models, whereas, in reality, water is mixed in the upper part of the reactor pressure vessel, presenting a natural averaging procedure. On the other hand, when relative errors are computed for per-primary loop parameters, values of the first three loops were used.

For the primary circuit (Tables \ref{tab:PerPLoop} and \ref{tab:Primary}), no relative error exceeds 3\%. For the secondary circuit (Table \ref{tab:PerSLoop}), errors are a bit larger, but never exceed 10\%. The larger error in the secondary circuit is due to the many geometrical approximations involved, and also due to the wide range of allowable parameter values at which secondary-circuit components can operate steadily and safely. For instance, the SG outlet mass flow rate has a lower limit of 407 kg/s, and an upper limit of 467 kg/s. While the SG water level has a plant value of 2.4 m, its lower limit is 1.7 m \cite{Groudev2000}.

Comparing the parameter values of Loop 4 and the other three loops, it can be observed that differences in the values are too small to have any statistical significance. It can be concluded that the input deck models VVER-1000 with sufficient accuracy for both geometry and parameter values. Having modeled both in-core fluid mixing and core inlet losses (cf. Section \ref{Sec:PV}), and having verified model fidelity in predicting in-core TH parameters, the input deck is now suitable for its intended use, i.e., predicting steady-state CHF values.  

It is worth mentioning that RELAP-SCDAPSIM contains a built-in model for CHF calculation that is based on the 1995 version of the Groeneveld LUT \cite{RELAPModels}. This model, however, is only defined for RELAP5 heat structures and cannot be implemented with our input deck which defines fission heat generation using the more accurate SCDAP heat structures. In addition, the 1995 version is obsolete compared to the 2006 version in terms of table values and correction factor correlations, and it isn't considered further in our analysis. 

\begin{table}[h!]
    \centering
    \footnotesize{
    \begin{tabular}[\textwidth]{l|ccccc|c|c}
    \hline
    Per-channel parameter & \multicolumn{5}{c|}{Calculated channel value} & Plant value & Error \\
     & 512 & 513 & 514 & 515 & 516 & & \\
    \hline
    Inlet coolant temperature [$^{\circ}$C]     & 296.9   & 295.5   & 295.9   & 296.4   & 297.0      & - & - \\  
    Outlet coolant temperature [$^{\circ}$C]    & 333.0   & 313.2   & 319.2   & 326.4   & 335.0      & 320 $\pm$ 3.5 & 1.7\% \\
    Coolant pressure [MPa]                      & 15.85   & 15.85   & 15.85   & 15.85   & 15.85      & - & - \\
    Axial flow velocity [m/s]                   & 6.202   & 5.895   & 6.222   & 6.635   & 6.299      & 7 & - \\
    Channel mass flux [kg/(m$^{2}$s)]           & 4,210   & 4,205   & 4,377   & 4,589   & 4,260      & - & - \\
    \hline
    \end{tabular}
    }
    \caption{Per channel parameters}
    \label{Tab:PerChannel}
\end{table}

\begin{table}[h!]
    \centering
    \footnotesize{
    \begin{tabular}{l|cccc|c|c}
    \hline
    Per-loop parameter & \multicolumn{4}{c|}{Calculated loop value} & Plant value & Error \\
     & Loop 1 & Loop 2 & Loop 3 & Loop 4 &  &  \\
    \hline
    SG inlet temperature [$^{\circ}$C]    & 323.8   & 323.8   & 323.8   & 323.8   & 318 $\pm$ 2 & 1.8\% \\
    SG outlet temperature [$^{\circ}$C]   & 294.6   & 294.6   & 294.6   & 294.5   & 287 $\pm$ 2 & 2.6\% \\
    Hot leg temperature [$^{\circ}$C]     & 323.8   & 323.8   & 323.8   & 323.8   & - & - \\
    Cold leg temperature [$^{\circ}$C]    & 294.8   & 294.8   & 294.8   & 294.7   & 287 $\pm$ 2 & 2.7\% \\
    SG pressure [MPa]                     & 15.71   & 15.71   & 15.71   & 15.71   & 15.64 & 0.4\% \\
    Hot leg mass flow rate [kg/s]         & 4,468   & 4,468   & 4,468   & 4,457   & - & - \\
    Cold leg mass flow rate [kg/s]        & 4,468   & 4,468   & 4,468   & 4,457   & - & - \\
    \hline
    \end{tabular}
    }
    \caption{Per primary loop parameters}
    \label{tab:PerPLoop}
\end{table}

\begin{table}[h!]
    \centering
    \footnotesize{
    \begin{tabular}{l|c|c|c}
    \hline
    Parameter & Calculated value & Plant value & Error \\
    \hline
    Reactor thermal power [MW]            & 3000    & 3000 & 0\% \\
    Pressurizer pressure [MPa]            & 15.72   & 15.65 & 0.4\% \\
    Pressurizer temperature [$^{\circ}$C] & 345.3   & 347 $\pm$ 1 & 0.5\% \\
    Coolant flow rate [kg/s]              & 17,861  & 17,610 $\pm$ 400 & 1.4\% \\
    \hline
    \end{tabular}
    }
    \caption{Parameters common to all primary loops}
    \label{tab:Primary}
\end{table}

\begin{table}[h!]
    \centering
    \footnotesize{
    \begin{tabular}{l|cccc|c|c}
    \hline
    Per-loop parameter & \multicolumn{4}{c|}{Calculated loop value} & Plant value & Error \\
     & Loop 1 & Loop 2 & Loop 3 & Loop 4 &  &  \\
    \hline
    SG inlet temperature [$^{\circ}$C]    & 209.7   & 209.7   & 209.7   & 209.7   & 220 $\pm$ 5 & 4.9\% \\
    SG outlet temperature [$^{\circ}$C]   & 281.4   & 281.4   & 281.4   & 281.4   & - & - \\
    SG pressure [MPa]                     & 6.566   & 6.566   & 6.566   & 6.566   & 6.17$-$6.56 & 3.2\% \\
    Main feedwater mass flow rate [kg/s]  & 400.2   & 400.2   & 400.2   & 399.5   & 437 $\pm$ 30 & 8.4\% \\  
    SG outlet mass flow rate [kg/s]       & 400.2   & 400.2   & 400.2   & 399.5   & 437 $\pm$ 30 & 8.4\% \\
    SG water level [m]                    & 2.170   & 2.170   & 2.170   & 2.170   & 2.4 $\pm$ 0.05 & 9.6\% \\
    \hline
    \end{tabular}
    }
    \caption{Per secondary loop parameters}
    \label{tab:PerSLoop}
\end{table}

\section{CHF prediction methods}

\subsection{Look-up tables}

\subsubsection{Groeneveld look-up table}
\label{GRO}

The Groeneveld CHF LUT is a set of normalized data points for a vertical 8-mm water-cooled round tube. It is based on 24,781 validated data points \cite{Groeneveld2007} and provides CHF values as a discrete function of pressure, $p$, mass flux, $G$, and equilibrium quality, $x$. Ranges of validity for the Groeneveld LUT, the Bobkov LUT as well as the empirical correlations are summarized in Table \ref{Tab:ValRanges}. For comparison purposes, VVER-1000 operational ranges can be found in Table \ref{ASD} of Section \ref{Sec:Results}.

\begin{table}[h!]
    \centering
    \scriptsize{
    \begin{tabular}{c|c|c|c|c|c}
    \hline 
		& Groeneveld LUT & Bobkov LUT & W-3 correlation & Biasi correlation & Bowring correlation \\
    \hline
Pressure, $p$ [MPa]						           & 0.1 to 21    & 0.11 to 20	  & 5.5 to 16     & 0.27 to 14                  & 0.2 to 19.0  \\  
Mass flow rate, $G$ [kg/(m$^{2}$s)]			       & 0 to 8000	  & 25 to 5000	  & -0.15 to 0.15 & 100 to 6000                 & 136 to 18600 \\
Equilibrium quality, $x$					       & -0.5 to 1	  & -0.5 to 1	  & 1356 to 6800  & $1/(1+\rho_f/\rho_g)$ to 1  & -            \\
Heated length, $L$ [m]							   & -		      & 0.8 to 7	  & 0.254 to 3.70 & 0.2 to 6                    & 0.15 to 3.7  \\
Rod diameter, $d$ [mm]						       & -			  & 5 to 13.5	  & -			  & -                           & -            \\
Hydraulic diameter, $D_{\mathrm{h}}$ [mm]          & -			  & 2.42 to 21	  & 15 to 18	  & 3 to 37.5                   & 2 to 45      \\
Relative rod pitch, $s/d$						   & -			  & 1.02 to 1.52  & -			  & -                           & -            \\
Inlet enthalpy, $h_{\mathrm{in}}$ [kJ/kg]  & -			  & -			  & $\geq$ 930.4  & -                           & -            \\
$\frac{\mathrm{Heated \ perimeter}}{\mathrm{Wetted \ perimeter}}$ & -  & -        &	0.88 to 1     & -  & -  \\
    \hline
    \end{tabular}
    }
    \caption{Ranges of validity for the CHF prediction methods \cite{Groeneveld2007, Bobkov2011, Todreas2012, Tong1996, Kumamaru1989, Kao1983}. Note that no validity ranges have been reported for the OKB-Gidropress correlation by Mozafari et al. \cite{Mozafari2013}.}
    \label{Tab:ValRanges}
\end{table}

The Groeneveld LUT is derived by \textit{(a)} statistically averaging CHF data points within each interval of pressures, mass fluxes, and qualities when experimental data are available, and \textit{(b)} extrapolation using known trends when data are not available \cite{Cheng2003, Groeneveld2007}. To apply the Groeneveld LUT to rod bundles, CHF values are first extracted from the table using a trilinear or tricubic \cite{Lekien2005} interpolation algorithm, and then are multiplied by several correction factors (Eq. \ref{Eq:KKK}). The formulas for these correction factors (Eqs. \ref{Eq:K1}-\ref{Eq:last}) are based on Refs. \cite{IAEA2001, Todreas2012}. As obvious in Table \ref{Tab:CHFFactors}, the correction factors reflect bundle-specific or subchannel-specific that are not incorporated in tube data \cite{IAEA2001}. Eq. \ref{Eq:KKK} assumes these effects are independent which is a first-order approximation. The accuracy of the Groeneveld LUT predictions for rod bundles depends to a large extent on the accuracy of the correction factor formulas. The most recent correction factor formulas are given in Eqs. \ref{Eq:K1}$-$\ref{Eq:K8}.

The effect of tube diameter on CHF is represented by $K_1$, where many values and correlations have been proposed for $n$. Originally, Groeneveld recommended using $n = 0.5$ and used this value in normalizing his LUT values \cite{IAEA2001, Groeneveld2007}. A value of $n = 1/3$ has been recommended by other researchers \cite{IAEA2001}. A recent analysis by Tanase et al. \cite{Tanase2009} has shown that the Wong correlation (Eq. \ref{Eq:Wong}) and Tanase correlation (represented by Table \ref{Tab:Tanase}) give the least RMS errors in predicting experimental round-tube data using the 2006 Groeneveld LUT. For subcooled conditions of interest in VVER rod bundles, Tanase et al. recommend using a value of $n$ = 0.25$-$0.33. Although correlations for $n$ have been based on round-tube data, they are extended to subchannels and other flow paths of irregular cross-sections for which the hydraulic diameter is used \cite{IAEA2001}. In this paper, the aforementioned approaches to defining $n$ will be assessed for use with VVER rod bundles.

\begin{equation}
\mathrm{CHF}_{\mathrm{bundle}} = K_1 \ K_2 \ K_3 \ K_4 \ K_5 \ K_6 \ K_7 \ K_8 \ \mathrm{CHF}_{\mathrm{table}}
\label{Eq:KKK}
\end{equation}

\begin{table}[h!]
    \centering
    \footnotesize{
    \begin{tabular}{ll}
    \hline
    Factor & Description \\
    \hline
    $K_1$ & Tube diameter factor \\
    $K_2$ & Bundle geometry factor \\
    $K_3$ & Mid-plane spacer factor for CANDU bundles \\
    $K_4$ & Heated length factor \\
    $K_5$ & Axial flux distribution factor \\
    $K_6$ & Radial flux distribution factor \\
    $K_7$ & Flow orientation factor \\
    $K_8$ & Vertical low flow factor \\
    \hline
    \end{tabular}
    }
    \caption{Correction factors for CHF values extracted from the LUT \cite{Todreas2012, IAEA2001}}
    \label{Tab:CHFFactors}
\end{table}

\begin{table}[h!]
    \centering
    \begin{tabular}{llllll}
    \hline
    Pressure (MPa) & Mass flux (kg/(m$^{2}$s)) & Quality \\
    \hline
    & & $-0.5$ to $-0.25$ & $-0.25$ to 0 & 0 to 0.5 & 0.5 to 1 \\
    \hline
    0.1$-$14 & 0$-$250 & $-0.2$ & $-0.2$ & $-0.2$ & $-0.3$ \\
    & 250$-$3000       & 0.4 & 0.4 & 0.5 & 0.6 \\
    & 3000$-$8000      & 0.3 & 0.3 & 0.4 & 0.4 \\
    14$-$21 & 0$-$250  & $-0.2$ & $-0.2$ & $-0.2$ & $-0.3$ \\
    & 250$-$3000       & 0.4 & 0.2 & 0.4 & 0.4 \\
    & 3000$-$8000      & 0.3 & 0.2 & 0.2 & 0.2 \\
    \hline
    \end{tabular}
    \caption{Tanase correlation for exponent $n$ \cite{Tanase2009}}
    \label{Tab:Tanase}
\end{table}

In the rod geometry factor, $K_2$, the absolute value does not exist in the original formula found in the literature \cite{IAEA2001, Todreas2012}, and has been introduced by the authors for convenience, otherwise, the value of $K_2$ would have been imaginary. The formula of $K_2$ is based on saturated qualities and hasn't been tested for negative subcooled qualities. This is likely because a pressurized test tube is harder to set up in the laboratory than a boiling test tube.

$K_3$, the spacer grid factor, is mainly developed for CANDU reactors where the local CHF enhancement can be as high as 150\% and has been tested for a limited number of other spacer geometries \cite {IAEA2001}. It has been recommended by IAEA-TECDOC-1203 \cite{IAEA2001} that for vertical rod bundles either $K_3$ or $K_4$ should be used. Because an independent measurement of the spacer grid pressure loss factor is not available only $K_4$ is used and $K_3 = 1$. Both $K_5$ and $K_6$ are 1 for subcooled conditions and are not employed either. $K_7$ = 1 for horizontal flows and $K_8$ = 1 for upward flow, so both of them are not used. The formula for $K_8$ is derived as recommended by IAEA-TECDOC-1203 \cite{IAEA2001} for $-400 < G < 0$, that is, to use linear interpolation between the table value for upward flow and CHF$_p$.

\begin{equation}
K_1 =
\left\{
\begin{array}{ll}
{\left[8/D_h\right]}^{n} & 2 \leq D_h \leq 25 \ \mathrm{mm} \\
0.57 & D_h>25 \ \mathrm{mm} \\
\end{array}
\right.
\label{Eq:K1}
\end{equation}

\begin{equation}
n = 0.58  \left[ 1 - 0.25 \ \mathrm{exp} \left(-2x\right) \right] \cdot \left( 1 - 15 \ D_h^{-6} \ G \right)
\label{Eq:Wong}
\end{equation}
where $D_h$ is the hydraulic diameter measured in mm.

\begin{equation}
K_2 =
\mathrm{min} \left[1, \left( 0.5 + \frac{2 (s-d)}{d} \right) \mathrm{exp} \left( -0.5 |x|^{1/3} \right) \right]
\label{Eq:K2}
\end{equation}

\begin{equation}
K_3 = 1 + A \ \mathrm{exp} \left[ - \frac{0.1 L_{s}}{D_h} \right]
\end{equation}

\begin{equation}
A = 1.5 \ \xi^{0.5} \left( 0.001 G \right)^{0.2}
\end{equation}
where $L_s$ and $\xi$ are the length and the pressure loss factor of the spacing device, respectively.

\begin{equation}
K_4 =
\left\{
\begin{array}{ll}
1    & L/D_h < 5 \\
\mathrm{exp}\left[e^{2 \alpha} \ L / D_h \right] & L/D_h \geq 5 
\end{array}
\right.
\label{Eq:K4}
\end{equation}

\begin{equation}
\alpha = \frac
{x \ \rho_f}
{x \ \rho_f + (1 - x) \ \rho_g}
\label{Eq:alpha1}
\end{equation}
where $L$ is the heated length from channel entrance to point of interest.

\begin{equation}
K_5 =
\left\{
\begin{array}{ll}
1 & x \leq 0 \\
q_{\mathrm{local}} / q_{\mathrm{BLA}} & x > 0 \\
\end{array}
\right.
\label{Eq:K5}
\end{equation}

\begin{equation}
q_{\mathrm{BLA}} = \frac{1}{L_B} \int_{z(x=0)}^{z_c} q(z) dz
\label{Eq:BLA}
\end{equation}
where $L_B$ is the boiling length $q_{\mathrm{BLA}}$ is the boiling-length average (BLA) flux.

\begin{equation}
K_6 =
\left\{
\begin{array}{ll}
1 & x \leq 0 \\
q(z)_{\mathrm{max}} / q(z)_{\mathrm{av}} & x > 0 \\
\end{array}
\right.
\label{Eq:K6}
\end{equation}

\begin{equation}
K_7 = 1 - \mathrm{exp} \left[ - \left( A / 3 \right)^{0.5} \right]
\end{equation}
\begin{equation}
A = \left( \frac{1-x}{1-\alpha} \right)^{2} \frac{f \ G^2}{g \ D_h \ \rho_f (\rho_f - \rho_g) \ \alpha^{0.5}}
\end{equation}
where $\alpha$ is the same defined in Eq. \ref{Eq:alpha}, $f$ is the channel's friction factor, and $g$ is the gravitational acceleration.

\begin{equation}
K_8 \ \mathrm{CHF}_{\mathrm{table}} =
\left\{
\begin{array}{ll}
2 \ \mathrm{CHF}_{p} - \mathrm{CHF}(|G|) & -400 < G < 0 \\
\mathrm{CHF}_{\mathrm{table}} & \mathrm{Otherwise} \\
\end{array}
\right.
\label{Eq:K8}
\end{equation}

\begin{equation}
\mathrm{CHF}_{p} = B \ (1-\alpha) \ \mathrm{CHF}(G=0, \ x=0)     
\end{equation}
\begin{equation}
B = 
\left\{
\begin{array}{ll}
1 & \alpha < 0.8 \\
\frac{0.8 + 0.2 \rho_f / \rho_g}{\alpha + (1-\alpha) \rho_f / \rho_g}  & \alpha \geq 0.8 \\
\end{array}
\right.
\label{Eq:last}
\end{equation}
where the minus sign refers to downward flow, $\mathrm{CHF}(|G|)$ is the CHF value at the upward flow, and $\mathrm{CHF}(G=0, \ x=0)$ is the CHF at pool boiling.

\subsubsection{Bobkov Look-up table}
\label{Bobkov}

The Bobkov LUT has been under continuous development for years. Its first version, the 1997 Bobkov LUT, comprises more than 4000 data points obtained from experiments on triangular assemblies performed at the Gidropress Experimental Design Office (OKB Gidropress), the State Scientific Center of the Russian Federation, the I.I. Leypunsky Institute Of Physics And Power Engineering (IPPE), as well as other organizations \cite{Bobkov2011}. The 2011 version of the Bobkov LUT is improved over the 1997 version in two aspects:
\begin{enumerate}
    \itemsep -0.3 em
    \item A considerable amount of new experimental data has been obtained in recent years, especially in poorly studied regions of low mass fluxes ($G <$ 500 kg/(m$^{2}$s)) and low pressures ($p <$ 3.0 MPa).
    \item The data have been made closer to the requirements of subchannel analysis by taking into account the influence of the peripheral zones of assemblies on the burnout phenomenon.
\end{enumerate}
The 2011 Bobkov LUT \cite{Bobkov2011} is a set of 5300 data points obtained on 49 experimental VVER rod bundles containing numbers of rods = 7, 9, 20, and 37. A ``slicing'' method has been applied to the Groeneveld LUT to identify outliers and assure LUT smoothness. However, no such method has been used during the derivation of the Bobkov LUT. In the diameter factor, $k_1$, a value of $n = 1/3$ is used, which is within the range recommend by Tanase et al. \cite{Tanase2009} for subcooled conditions. The thermal diameter, $D_t$ appears in the original formulation of the Bobkov correction factors; however, we assume the thermal and hydraulic diameters to be equivalent, i.e., the ratio of the heated and wetted perimeters = 1. $k_2$ corrects for the relative rod pitch, and is nearly analogous to Groneveld's $K_2$ except that $k_2$ depends only on geometrical parameters, while a quality dependence is included in Groeneveld's factor (Eq. \ref{Eq:K2}). This is likely because subcooled qualities do not change significantly along a heated channel as opposed to saturated qualities on which Groeneveld factors are based as argued previously. For VVER-1000, $s = 12.75$ mm, and $d = 9.1$ mm \cite{Tabadar2018}.

$k_3$ accounts for the heated length effect and is analogous to Groeneveld's $K_4$. As with the bundle geometry effect, Bobkov's factor depends only on geometry, whereas Groeneveld's factor is a function of both geometry, and a hydrodynamic parameter, $\alpha$ (Eq. \ref{Eq:alpha1}). In the Groeneveld LUT, the effect of the spacer grid geometry is incorporated in $K_2$, whereas CHF enhancement due to spacer grids is only accounted for CANDUs $K_4$ \cite{IAEA2001}. In the Bobkov LUT, this effect is accounted for by $k_4$. An independent measurement of the VVER spacer gird friction factor ($\xi$ of Eq. \ref{KK4}) is, however, not available, and $k_4$ cannot be used. An alternative and more practical approach to account for the spacer grid effect in VVER rod bundles has been proposed by Bolshakov et al. \cite{Bolshakov2011}. Bolshakov et al. proposed a spacer grid factor of the form given by Eq. \ref{Eq:FB}.

\begin{equation}
F_B = 1 + a_1 F_1 \left( \xi \right) F_2 \left( z_s \right)  F_3 \left( G \right) F_4 \left( x \right) = 1 + a_2 \mathrm{exp} \left[ - (z_s/(bD_h))^m \right] (G/1000)^n (c+d|x|)
\label{Eq:FB}
\end{equation}
where $a_2$, $b$, $c$, $d$, $m$, and $n$ are empirical constants. $F_1 \left( \xi \right)$ has been absorbed in $a_1$ because it is constant for each configuration of spacer grids and the difficulty of determining the spacer grid friction factor is eliminated. Because the form of $F_4 \left( x \right)$ is based on measurements for saturated qualities \cite{Bolshakov2011}, the absolute value of $x$ was used. For VVER rod bundles, $a_2=1.6$, $b=7.6$, $c=0.2$, $d=1$, $m=1$, and $n=1$ \cite{Bolshakov2011}. In this study, $F_B$ is used with the Bobkov LUT instead of $k_4$.

A phenomenon that arises in CHF experiments on rod bundles is the unbalance between heating in central and peripheral subchannels \cite{Bobkov2011}. Smaller mass fluxes at the bundle boundaries cause the occurrence of CHF at TH conditions that are not representative of a full-scale VVER assembly. This effect has been confirmed by Kao and Kazimi \cite{Kao1983} using a subchannel analysis of a 9-rod BWR bundle. Bobkov et al. \cite{Bobkov2011} have observed that the smaller the number of rods in the experimental bundle, the more pronounced the effect. To account for this effect to a first-order approximation, a correction factor, $k_5$, is introduced which is defined as the ratio between the thermal diameter of the assembly's ``central subchannels,'' $D_t$, to the thermal diameter of the whole assembly, $D_{ta}$. This reasoning is based on the enthalpy imbalance approach. The degree to which the assemblies used for constructing the LUT were ``unbalanced" varied from 0.42 to 0.93 \cite{Bobkov2011}; the value 0.93 corresponding to a bundle containing 37 rods. The number of fuel rods in a VVER-1000 assembly is 311 \cite{Tabadar2018}, so we can safely assume that $k_5$ = 1.

\begin{equation}
\mathrm{CHF}_{\mathrm{bundle}} = k_1 \ k_2 \ k_3 \ k_4 \ k_5 \ \mathrm{CHF}_{\mathrm{table}}
\label{Eq:Bobkov}
\end{equation}

\begin{equation}
k_1 = \left[ 9.36 / D_t \right]^{1/3}
\end{equation}
where $D_t$ is measured in mm.

\begin{equation}
k_2 =
\left\{
\begin{array}{ll}
0.82 - 0.7 \ \mathrm{exp} \left[ -35 (s/d - 1) \right] & s/d \leq 1.1 \\
0.2 + 0.57 s/d & 1.1 < s/d < 1.52 \\
\end{array}
\right.
\end{equation}

\begin{equation}
k_3 = 1 + 0.6 \ \mathrm{exp} \left( -0.01 L/D_t \right)
\label{Eq:HL}
\end{equation}

\begin{equation}
k_4 = 1 + A \ \mathrm{exp} \left( -0.1 z_s / D_t \right)
\label{ASD2}
\end{equation}

\begin{equation}
A = 1.5 \ \xi^{0.5} \ \left( G/1000 \right)^{0.2}
\label{KK4}
\end{equation}
where $z_s$ is the distance from coolant outlet to the nearest spacer grid, and $\xi$ is the spacer grid friction factor \cite{Mozafari2013}.

\begin{equation}
k_5 = D_t / D_{ta}
\label{k5}
\end{equation}

The Bobkov LUT is used as the benchmark against which other CHF predictors are compared for the following reasons:
\begin{enumerate}
    \itemsep -0.3 em
	\item Assessing the uncertainty of the Bobkov LUT (along with its correction factors) in predicting experimental CHF values in VVER rod bundles, Bobkov et al. \cite{Bobkov2011} report that for experiments based on local conditions, the mean error in the dataset is near zero, and the RMS error is less than 15\%. For experiments based on constant inlet conditions, the mean error in the dataset is also near zero, whereas the RMS error is less than 8\% \cite{Bobkov2011}.      
    \item The Bobkov is based on experiments conducted directly using VVER rod bundles, and thus any specifics of the internal channel design (e.g., the effect of channel's convexity on the liquid film distribution) needn't be accounted for.
    \item The experimental datasets on which the Bobkov LUT is based have undergone a sufficient amount of scrutiny to ensure their accuracy. For instance, Groeneveld et al. \cite{Groeneveld1996} have examined the IPPE dataset and reported that it satisfied the heat balance criterion (i.e., power approximately equals mass flow rate $\times$ specific enthalpy\footnote{When practically testing heat balance, the measured power is the power input to the electric heaters used to heat the tubes inside which water flows \cite{Pioro2000}. Power loss naturally exists and the product of mass flow rate and specific enthalpy is usually $\sim$1\% \cite{Pioro2000} less than the measured power.}). Also, Kirrilov \cite{Cheng2003} compared the IPPE dataset to CHF data obtained in PWR rod bundles and concluded that ``there is no significant difference'' between the two datasets.
\end{enumerate}

\subsubsection{Trilinear interpolation}
\label{INTERPOL}

\begin{figure}[h]
\centering
\includegraphics[width=0.3\textwidth]{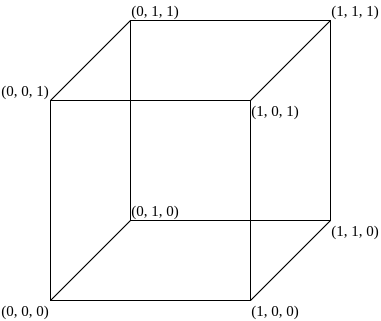}
\caption{Cube for the Bourke algorithm (reproduced from Ref. \cite{Bourke1997})}
\label{Fig:Cube}
\end{figure}

Because LUT entries are discrete, a trilinear or tricubic (e.g., that of Lekien and Marsden \cite{Lekien2005}) interpolation algorithm needs to be employed. In this paper, the time complexity of two trilinear interpolation algorithms is assessed:
\begin{enumerate}
\itemsep -0.3 em
\item Standard linear interpolation extended to three dimensions. For 3 variables, there are $3!=6$ possible arrangements for performing a standard linear interpolation. The algorithm for the $G$-$x$-$p$ arrangement is presented in Eqs. \ref{A1}$-$\ref{A2}.
\item The Bourke interpolation algorithm \cite{Bourke1997}. The Bourke algorithm normalizes the bounding values of the three variables, constructing a cube of unit length (Figure \ref{Fig:Cube}), and then performs the interpolation by applying the geometrically inspired formula, Eq. \ref{Eq:Bourke}, where $a$, $b$, and $c$ are normalized parameters.
\end{enumerate}
The Bourke algorithm is symmetric for the 3 interpolated variables and is thus simpler to execute. The standard algorithms, however, contain 36 floating-point operations (FLOPs) per iteration, compared to 49 FLOPs per iteration for the Bourke algorithm. Two approaches exist for determining the time complexity (or time cost) of an algorithm \cite{Kube2013}:
\begin{enumerate}
    \itemsep -0.3 em
    \item Analyze the asymptotic behavior of the algorithm by counting the statements executed and writing the time cost function as a big-O notation, e.g., O($n^2$), where $n$ is the number of inputs to the algorithm.
    \item Implement the algorithm, run it and measure the elapsed time.
\end{enumerate}
The problem with asymptotic algorithm analysis is that two algorithms may belong to the same class of time cost function, i.e., have the same argument inside the big-O, but in reality, take very different times to solve a problem of the same size $n$ \cite{Kube2013}. That's why time measurement is more reliable in assessing algorithms. Because the algorithm you are trying to ``time'' isn't the only program running on your computer, measured times usually contain noise. That's why instead of the elapsed time of one run, the averages and standard deviations of many runs are presented.

\begin{equation}
\mathrm{CHF}_{1} = \frac{G - G_0}{G_1 - G_0} \cdot (\mathrm{CHF}_{100} - \mathrm{CHF}_{000}) + \mathrm{CHF}_{000}
\label{A1}
\end{equation}

\begin{equation}
\mathrm{CHF}_{2} = \frac{G - G_0}{G_1 - G_0} \cdot (\mathrm{CHF}_{110} - \mathrm{CHF}_{010}) + \mathrm{CHF}_{010}    
\end{equation}

\begin{equation}
\mathrm{CHF}_{3} = \frac{x - x_0}{x_1 - x_0} \cdot (\mathrm{CHF}_{2} - \mathrm{CHF}_{1}) + \mathrm{CHF}_{1}
\end{equation}

\begin{equation}
\mathrm{CHF}_{4} = \frac{G - G_0}{G_1 - G_0} \cdot (\mathrm{CHF}_{101} - \mathrm{CHF}_{001}) + \mathrm{CHF}_{001}    
\end{equation}

\begin{equation}
\mathrm{CHF}_{5} = \frac{G - G_0}{G_1 - G_0} \cdot (\mathrm{CHF}_{111} - \mathrm{CHF}_{011}) + \mathrm{CHF}_{011}    
\end{equation}

\begin{equation}
\mathrm{CHF}_{6} = \frac{x - x_0}{x_1 - x_0} \cdot (\mathrm{CHF}_{5} - \mathrm{CHF}_{4}) + \mathrm{CHF}_{4}
\end{equation}

\begin{equation}
\mathrm{CHF}_{Gxp} = \frac{p - p_0}{p_1 - p_0} \cdot (\mathrm{CHF}_{6} - \mathrm{CHF}_{3}) + \mathrm{CHF}_{3}
\label{A2}
\end{equation}

\begin{equation}
\begin{array}{rc}
\mathrm{CHF}_{abc} = & \mathrm{CHF}_{000} \ (1 - a) \ (1 - b) \ (1 - c) + \mathrm{CHF}_{100} \ a \ (1 - b) \ (1 - c) \\
& + \ \mathrm{CHF}_{010} \ (1 - a) \ b \ (1 - c) + \mathrm{CHF}_{001} \ (1 - a) \ (1 - b) \ c \\
& + \ \mathrm{CHF}_{101} \ a \ (1 - b) \ c + \mathrm{CHF}_{011} \ (1 - a) \ b \ c \\
& + \ \mathrm{CHF}_{110} \ a \ b \ (1 - c) + \mathrm{CHF}_{111} \ a \ b \ c
\end{array}
\label{Eq:Bourke}
\end{equation}

\subsection{Empirical correlations}

\subsubsection{W-3 correlation}

It is the most widely available correlation for evaluating CHF in PWRs. The W-3 correlation can be applied to round, rectangular, and rod-bundle flow geometries. It has been developed for axially uniform heat flux, with a correcting factor for non-uniform flux distribution \cite{Todreas2012}, the Tong's factor, discussed in detail in Section \ref{CHFTrend}. As concluded later in Section \ref{Sec:Tong2}, non-uniform heat flux doesn't affect the CHF in subcooled conditions, and the uniform version of the W-3 correlation is used, which is defined in SI units by Eq. \ref{Eq:W3} \cite{Todreas2012, Tong1996, Cheng2003}.

\begin{equation}
\begin{array}{rc}
\mathrm{CHF}_{\mathrm{W-3}} = & \left\{ \left( 2.022 - 0.06238  p\right)+ \left( 0.1722 - 0.01427  p \right)  \mathrm{exp} \left[ \left( 18.177 - 0.5987   p \right)   x \right] \right\} \\
 & \times \left[ 2.326 ( 0.1484 - 1.596   x ) + 0.1729   x  |x|)  G + 3271 \right] \\
& \times \left[ 1.157 - 0.869  x \right] \times  \left[ 0.2664 + 0.8357   \mathrm{exp} \left( -124.1   D_h \right) \right] \\
& \times \left[ 0.8258 + 0.0003413 \left( h_{\mathrm{sat}} - h_{\mathrm{in}} \right) \right]
\end{array}
\label{Eq:W3}
\end{equation}

$\mathrm{CHF}_{\mathrm{W-3}}$ is measured in kW/m$^2$, $p$ in MPa, $G$ in kg/(m$^{2}$s), $D_h$ in m, $h_{\mathrm{sat}}$ (saturated liquid enthalpy) and $h_{\mathrm{in}}$ (channel inlet enthalpy) in kJ/kg.


\subsubsection{Biasi correlation}

The Biasi correlation had been used as the standard method of calculating the CHF before the era of LUTs and had been implemented in the early versions of nuclear codes, e.g., RELAP5 MOD 2, TRAC-PF1 MOD 2, and TRAC-BD1 \cite{RELAPModels, TRACE, Kumamaru1989}. It is defined by Eq. \ref{Eq:Biasi} \cite{Kumamaru1989}.

\begin{equation}
\mathrm{CHF}_{\mathrm{Biasi}} = \mathrm{max} \left[ \frac{1.883 \times 10^4} {D_{h}^{\alpha} \times G^{1/6}} \left( \frac{A}{G^{1/6}} - x \right), \ \frac{3.78 \times 10^4 \times B}{D_{h}^{\alpha} \times G^{0.6}} \left(1 - x \right) \right]
\label{Eq:Biasi}
\end{equation}

where $\alpha$ is defined by Eq. \ref{Eq:alpha}, $A$ is defined by Eq. \ref{Eq:y} and $B$ is defined by Eq. \ref{Eq:h}.
\begin{equation}
\alpha =
\left\{
\begin{array}{ll}
0.6  & D_h < 1 \ \mathrm{cm} \\
0.4  & D_h \geq 1 \ \mathrm{cm}
\end{array}
\right.
\label{Eq:alpha}
\end{equation}

\begin{equation}
A = 0.7249 + 0.099 \ p \times \mathrm{exp} \left(-0.032 \ p \right)
\label{Eq:y}
\end{equation}

\begin{equation}
B = -1.159 + 0.149 \ p \times \mathrm{exp} \left( -0.019 \ p \right) + \frac{8.99 \ p} {10 + p^2}
\label{Eq:h}
\end{equation}

In the equations of the Biasi correlation (Eqs. \ref{Eq:Biasi}$-$\ref{Eq:h}), $D_h$ is measured in cm, $G$ in g/cm$^2 \cdot \mathrm{s}$, $p$ in atm, and $\mathrm{CHF}_{\mathrm{Biasi}}$ in kW/m$^2$.

\subsubsection{Bowring correlation}
The Bowring correlation is based on 3800 experimental data points, and its reported RMS error is 7\% \cite{Kolev2011}. In SI units, it is described by Eqs. \ref{Bow1}$-$\ref{Bow2} \cite{Todreas2012}.

\begin{equation}
\mathrm{CHF}_{\mathrm{Bowring}} = 0.001 \frac{A-B \ h_{\mathrm{fg}} \ x}{C}
\label{Bow1}
\end{equation}

\begin{equation}
A = \frac{0.57925 \ F_1 \ h_{\mathrm{fg}} \ D_h \ G}{1 + 0.0143 \ F_2 \ D_h^{0.5} \ G}
\end{equation}

\begin{equation}
B = 0.25 \ D_h \ G
\end{equation}

\begin{equation}
C = \frac{0.077 \ F_3 \ D_h \ G}{1 + 0.347 \ F_4 \ \left( G / 1356 \right)^n}
\end{equation}

\begin{equation}
p_r = 0.145 p
\end{equation}

\begin{equation}
n = 2-0.5 \ p_r
\end{equation}

For $p_r < 1$:

\begin{equation}
\begin{array}{c}
F_1 = \{ p_r^{18.942} \ \mathrm{exp} \left[ 20.89 (1-p_r) \right] + 0.917 \} / 1.917 \\
F_2 = F_1 / \left( \{p_r^{1.316} \ \mathrm{exp} \left[ 2.444 \left( 1 - p_r \right) \right] + 0.309 \} / 1.309 \right) \\
F_3 = \{ p_r^{17.023} \ \mathrm{exp} \left[  16.658 (1-p_r) \right] + 0.667 \} / 1.667 \\
F_4 = F_3 \ p_r^{1.649}
\end{array}
\end{equation}

For $p_r \geq 1$:

\begin{equation}
\begin{array}{c}
F_1 = p_r^{-0.368} \mathrm{exp} \left[ 0.648 \left( 1 - p_r \right) \right]
\\
F_2 =  F_1 / \{ p_r^{-0.448} \mathrm{exp} \left[ 0.245 \left( 1 - p_r \right) \right] \} \\
F_3 = p_r^{0.219} \\
F_4 = F_3 \ p_r^{1.649}
\end{array}
\label{Bow2}
\end{equation}

$\mathrm{CHF}_{\mathrm{Bowring}}$ is measured in kW/m$^2$, $h_{\mathrm{fg}}$ in J/kg, $D_h$ in m, $G$ in kg/(m$^{2}$s), and $p$ in MPa \cite{Kolev2011}.

\subsubsection{OKB-Gidropress correlation}

As reported by Mozafari et al. \cite{Mozafari2013}, the OKB-Gidropress correlation is found in the FSAR of Unit 1 of BNPP. It is based on experimental studies performed on VVER rod bundles by OKB-Gidropress, VVER designer, and is defined for uniform heat distribution by Eq. \ref{Eq:VVER}.

\begin{equation}
\mathrm{CHF}_{\mathrm{OKB}} = 795 \left(1 - x\right)^{0.105 p - 0.5} \times G^{0.184 - 0.311  x} \times \left(1 - 0.0185  p\right)
\label{Eq:VVER}
\end{equation}
${\mathrm{CHF}}_{\mathrm{OKB}}$ is measured in kW/m$^2$, $p$ in MPa, and $G$ in kg/(m$^{2}$s).

\subsection{CHF trend}
\label{CHFTrend}

Generally, the CHF is a decaying function in the $q$-$z$ space due to its dependence on upstream conditions, which is caused by two independent physical effects: \textit{(a)} the heated length effect, and \textit{(b)} the non-uniform axial heat flux effect (i.e., the so-called memory effect). Heated length factors have been exclusively used with LUTs as discussed in Sections \ref{GRO} and \ref{Bobkov}. There exist four approaches to take account of the memory effect:
\begin{enumerate}
\itemsep -0.3 em

\item The overall power approach, which assumes the critical power is the same independent of the heat flux profile, provided that channel geometry and inlet conditions are fixed.

\item The local conditions approach, which assumes the CHF is entirely a local phenomenon. The non-uniform axial heat flux factor, $F_{L}$, is determined by performing power increase iterations either experimentally or numerically (cf. Section \ref{intro}) on the same channel for both uniform and non-uniform heat flux profiles and taking their ratio (Eq. \ref{Eq:FL}).

\item Tong's approach, derived by Tong et al. \cite{Yang2006}, also uses a non-uniform axial heat flux factor, $F_{T}$, defines it by Eq. \ref{Eq:TongApp} \cite{Todreas2012}, and then calculates the non-uniform CHF by Eq. \ref{Eq:TongF}. This approach has only been applied to the W-3 correlation \cite{Todreas2012}.

\item The BLA approach, proposed by Groeneveld \cite{Yang2006} for use with the Groeneveld LUT, defines a BLA heat flux (Eq. \ref{Eq:BLA} of Section \ref{GRO}), and a non-uniform axial heat flux factor, $K_5$ (Eq. \ref{Eq:K5} of Section \ref{GRO}). The non-uniform CHF is then calculated by multiplying $K_5$ by the uniform CHF. This approach is only defined for saturated qualities, while Tong's approach can be applied to both saturated and subcooled conditions.
\end{enumerate}

\begin{equation}
F_{L} = \frac{\mathrm{CHF}_{\mathrm{uniform}}}{\mathrm{CHF}_{\mathrm{non-uniform}}}
\label{Eq:FL}
\end{equation}

\begin{equation}
F_{T} = \frac{c e^{-cz}}{q(z) \ (1-e^{-cz})} \ \int_{0}^{z} q(z') \ e^{cz'} \ dz'
\label{Eq:TongApp}
\end{equation}

\begin{equation}
c = 185.6 \ \frac{\left[ 1 - x(z) \right]^{4.31}}{G^{0.478}}
\label{Eq:c}
\end{equation}

\begin{equation}
\mathrm{CHF}_{\mathrm{non-uniform}} = \frac{\mathrm{CHF}_{\mathrm{uniform}}}{F_{T}}
\label{Eq:TongF}
\end{equation}

For subcooled qualities, the value of $c$ (Eq. \ref{Eq:c}) is large, and thus the memory effect on the CHF is small. Additionally, based on the liquid-sublayer dryout model \cite{Cheng2003} that attempts to phenomenologically explain the DNB mechanism, it has been argued by Yang et al. \cite{Yang2006} that sublayer evaporation at high subcooling is mainly caused by local overheating, and the dependence of the CHF on upstream conditions is small. Bubble-layer models \cite{Todreas2012} of DNB argue that the boiling crisis depends on the ``history'' of bubble layer evolution upstream, asserting that DNB is not a completely local phenomenon. On contrary, Cheng and M{\"{u}}ller \cite{Cheng2003} have compared the predictions of both the near-wall bubble-crowding model (which is a type of bubble-layer models) and the liquid-sublayer dryout model with five round-tube experimental data points. They have found that the near-wall bubble-crowding model underpredicts the CHF by up to about 25\%, whereas the liquid-sublayer dryout model overpredicts the CHF by up to about 30\%. However, no reliable conclusion can be drawn from these very few data points. Therefore, many researchers (e.g., Bobkov et al. \cite{Bobkov2011}, Groeneveld et al. \cite{Groeneveld2007}, and Yang et al. \cite{Yang2006}) have neglected the CHF dependence on upstream history in subcooled conditions and treated it as a solely local phenomenon. The validity of this assumption is evaluated in this study.

\subsection{Error estimation}

There exist four error estimators to compare $n$ discrete model predictions, $P_i$, to $n$ actual observations, $O_i$ \cite{Willmott2005, Taebi2017}:
\begin{enumerate}
    \itemsep -0.3 em
    \item Mean Absolute Error (MAE) (Eq. \ref{Eq:MAE}),
    \item Root Mean Square Error (RMSE) (Eq. \ref{Eq:RMSE}),
    \item Normalized Mean Absolute Error (NMAE), and
    \item Normalized Root Mean Square Error (NRMSE),
\end{enumerate}
where the NMAE and NRMSE are normalized by the mean value of the actual observations, $\bar{O}$. In general, there is no agreement over the exact formulas for MAE and RMSE, nor over the exact statistical parameter by which NMAE or NRMSE are normalized. Some authors (e.g., Knoll \cite{Knoll2010}) divide by $n-1$ instead of $n$ in Eqs. \ref{Eq:MAE} and \ref{Eq:RMSE}. Others (e.g., Surridge et al. \cite{Surridge2014}) normalize the NMAE and NRMSE by the interquartile range. NRMSE is often reported in thermal hydraulics literature as the RMS error, although no explicit mention exists of the exact formula used for the NRMSE, which impedes the ability to compare results reported in different research papers. In addition, Willmott and Matsuura \cite{Willmott2005} have proved that the NRMSE is not a measure of model uncertainty but, rather, is a measure of error distribution among the $n$ points. The more error is ``concentrated'' within a fewer number of points, the larger the NRMSE, which is not representative of the performance of the whole model. That is, the NRMSE gives more weight to outliers. In this study, the NMAE and NRMSE are used to assess the performance of the CHF predictors, where MAE and RMSE are calculated by Eqs. \ref{Eq:MAE} and \ref{Eq:RMSE} respectively, and both are normalized by the mean value of CHF along the channel as calculated by the Bobkov LUT. The NRMSE is only calculated for completeness.

\begin{equation}
\mathrm{MAE} = \frac{\sum_{i=1}^{n} |O_i-P_i|}{n}
\label{Eq:MAE}
\end{equation}

\begin{equation}
\mathrm{RMSE} = \left[\frac{\sum_{i=1}^{n} \left(O_i-P_i\right)^{2}}{n}\right]^{1/2}
\label{Eq:RMSE}
\end{equation}

\section{Results}
\label{Sec:Results}
A steady-state simulation of VVER-1000 was performed on RELAP-SCDAPSIM for 1500 seconds at 112\% of nominal power. The resulting parameter values of the hot channel are presented in Table \ref{ASD}.

\begin{table}[h!]
    \centering
    \footnotesize{
    \begin{tabular}{c|c|c|c|c}
    \hline 
    Node & $p$ (MPa) & $G$ (kg/(m$^{2}$s)) & $x$ & $\alpha$ (Eq. \ref{Eq:alpha1}) \\
    \hline
    1  & 15.57 &     4514 &	$-$0.4283 & 1.792 \\
    2  & 15.57 &	 4505 &	$-$0.4271 & 1.794 \\
    3  & 15.56 &	 4499 &	$-$0.4259 & 1.797 \\ 
    4  & 15.55 &	 4494 &	$-$0.4247 & 1.799 \\
    5  & 15.54 &	 4491 &	$-$0.4236 & 1.801 \\ 
    6  & 15.53 &	 4489 &	$-$0.4225 & 1.803 \\ 
    7  & 15.52 &	 4488 &	$-$0.4213 & 1.805 \\
    8  & 15.51 &	 4487 &	$-$0.4202 & 1.807 \\ 
    9  & 15.51 &	 4487 &	$-$0.4191 & 1.809 \\ 
    10 & 15.50 &	 4484 &	$-$0.4181 & 1.811 \\
    \hline
    \end{tabular}
    }
    \caption{Pressure, mass flux, equilibrium quality, and $\alpha$ (Eq. \ref{Eq:alpha1}) calculated for the hot channel nodes}
    \label{ASD}
\end{table}

\subsection{Memory effect}
\label{Sec:Tong2}

Tong's $F_T$ has been calculated in the subcooled conditions of VVER-1000 (Table \ref{ASD}) for \textit{(a)} the profile of Table \ref{Tab:APF}, and \textit{(b)} a $q$ profile sampled from a sine function. The results of such calculation are presented in Table \ref{Tab:TongF}. Because APFs rather than actual heat fluxes were used, only relative values of $F_T$ are meaningful, and the values in Table \ref{Tab:TongF} have been normalized by the values of $F_T$ for a uniform profile. $F_T$ approximately equals 1 for both profiles at all nodes. It can be concluded that in subcooled conditions, the CHF is insensitive to heat flux non-uniformity, and only the effect of heated length gives the CHF its decaying trend. That is, a heated length factor should be used even with empirical correlations. In the rest of the paper, no memory effect factor is used, and the Bobkov heated length factor, $k_3$ of Eq. \ref{Eq:HL}, is used with the four empirical correlations assessed. This conclusion agrees with assertions made by Bobkov et al. \cite{Bobkov2011}, Groeneveld et al. \cite{Groeneveld2007}, and Yang et al. \cite{Yang2006}, and contradict the results of Cheng and M\"{u}ller \cite{Cheng2003}. $k_3$ was chosen because it is specifically formulated for VVER rod bundles.

\begin{table}[h!]
    \centering
    \begin{tabular}{c|cccccccccc}
    \hline
    Node & 1 & 2 & 3 & 4 & 5 & 6 & 7 & 8 & 9 & 10 \\
    \hline
    Table \ref{Tab:APF} & 1.0000 & 0.9979 & 0.9999 & 1.0002 & 1.0001 & 1.0000 & 1.0000 & 1.0000 & 1.0003 & 1.0028 \\
    Sine-sampled & 1.0000 & 0.9978 & 0.9985 & 0.9988 & 0.9991 & 0.9992 & 0.9993 & 0.9994 & 0.9995 & 0.9996 \\
    \hline
    \end{tabular}
    \caption{Tong's $F_T$ at subcooled conditions for two heat flux profiles}
    \label{Tab:TongF}
\end{table}

\subsection{Diameter factor exponent}

\begin{figure}[h!]
    \centering
    \includegraphics[width=\textwidth]{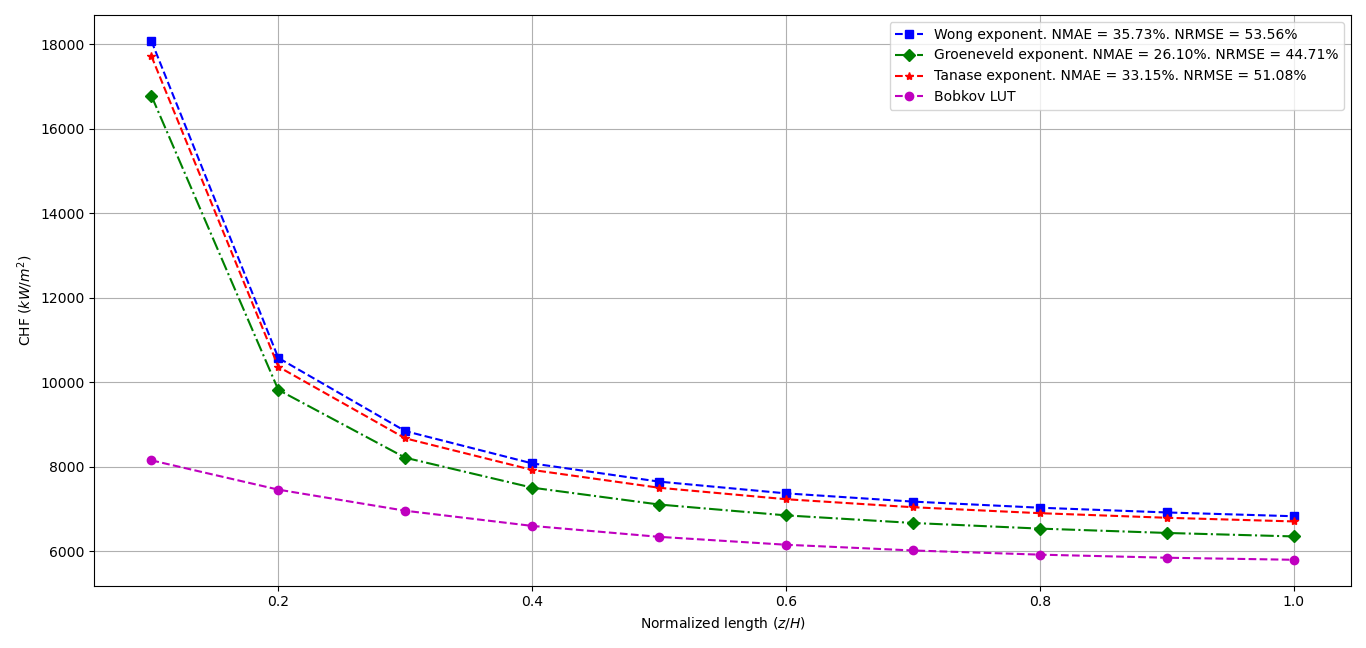}
    \caption{Diameter factor exponents of Wong, Groeneveld, and Tanase applied to the Groeneveld LUT and the resulting CHF values compared to the Bobkov LUT}
    \label{Fig:CHF-Exp}
\end{figure}

Because the Groeneveld LUT is widely used \cite{Groeneveld2007} and its 1995 version has been integrated into many codes, e.g., RELAP5 MOD3.3 \cite{RELAPModels} and TRACE V5.0 \cite{TRACE}, it is of interest to study its predictability of the CHF in VVER rod bundles. For this purpose, Diameter factor exponents of Wong, Groeneveld, and Tanase have been applied to the Groeneveld LUT, and the resulting CHF values have been compared to the Bobkov LUT. The results of such comparison are illustrated in Figure \ref{Fig:CHF-Exp}, and the values of the Wong exponent along the channel are shown in Table \ref{Tab:Wongn}. Although both Tanase and Wong correlations give better performance over the whole range of parameters covered by the Groeneveld LUT, it has been found that the traditional Groeneveld exponent $n=1/2$ gives better predictability of the CHF in VVER conditions. It may be concluded from this counter-intuitive result that the diameter effect in round tubes will not be extrapolated to rod bundles and that further research is needed to quantify the diameter effect on rod bundles, especially in reactors of unusual channel geometries, e.g., advanced reactors.

\begin{table}[h!]
    \centering
    \begin{tabular}{c|cccccccccc}
    \hline
    Node & 1 & 2 & 3 & 4 & 5 & 6 & 7 & 8 & 9 & 10 \\
    \hline
    $n$ & 0.2265 & 0.2273 & 0.2280 & 0.2288 & 0.2295 & 0.2303 & 0.2310 & 0.2317 & 0.2325 & 0.2331 \\
    \hline
    \end{tabular}
    \caption{Wong's exponent, $n$, for the hot channel of VVER-1000}
    \label{Tab:Wongn}
\end{table}

\subsection{Assessment of CHF predictors}

\begin{figure}[h!]
    \centering
    \includegraphics[width=\textwidth]{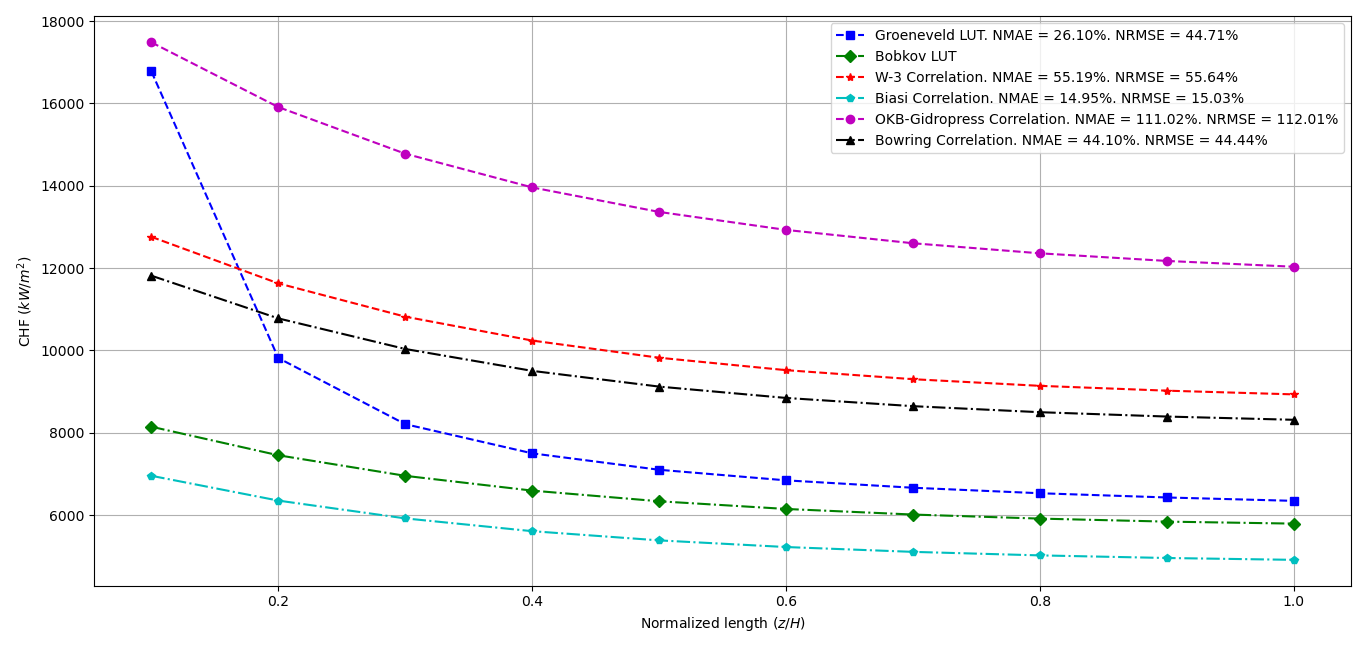}
    \caption{Comparison of CHF predictors for VVER rod bundles}
    \label{Fig:CHF-Predictors}
\end{figure}

\begin{figure}[h!]
    \centering
    \includegraphics[width=\textwidth]{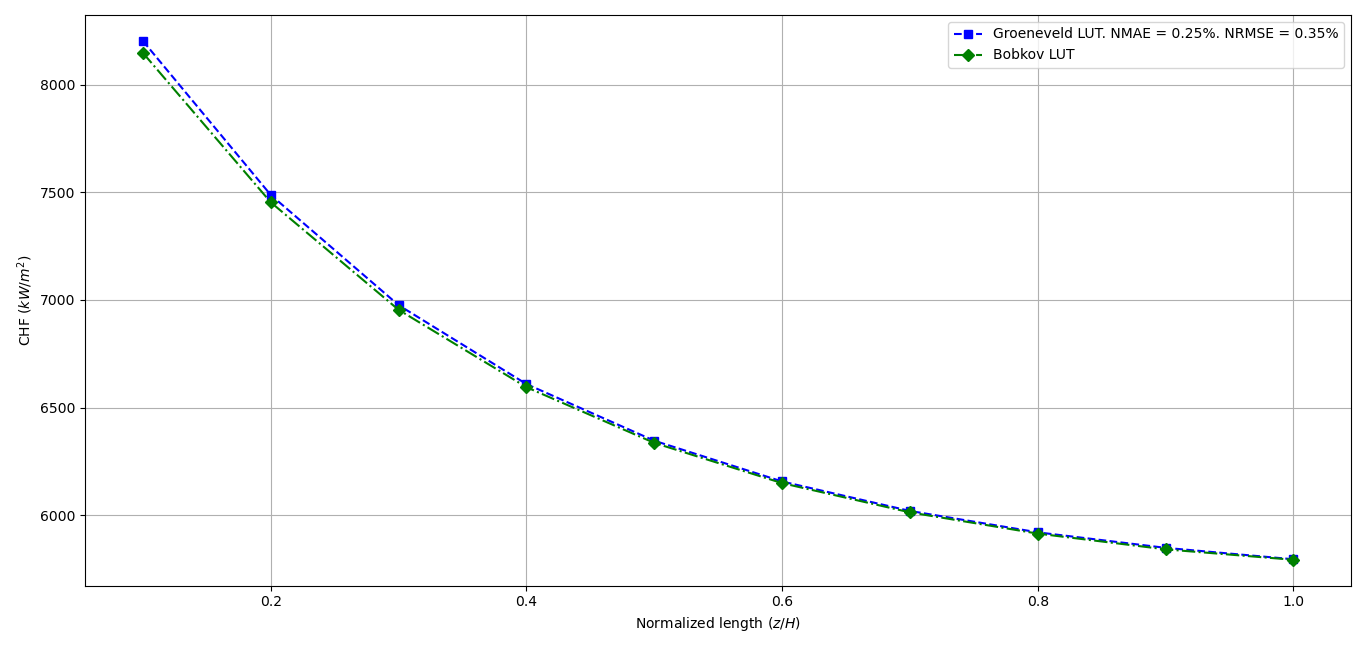}
    \caption{Applying the Bobkov heated length factor, $k_3$, to the Groeneveld LUT}
    \label{Fig:GB-LUT}
\end{figure}

Four CHF predictors have been used to calculate the CHF in the hot channel of VVER-1000, and the results have been compared to the Bobkov LUT CHF values (Figure \ref{Fig:CHF-Predictors}). A few observations about these results are in order.

\begin{itemize}

\itemsep -0.3 em

\item Upon comparison to the Bobkov LUT, the Groeneveld LUT curve exhibits the best performance among all of the CHF predictors with an NMAE = 26.10\%. At the channel inlet, the Groeneveld LUT overestimates the CHF by more than a factor of 2 and exhibits a very steep descent, which explains the large value of the NMRSE = 44.71\% (Remember that the NRMSE gives more weight to outliers). Downstream, it approaches the Bobkov LUT curve. This steep gradient at the channel inlet is mainly due to Groeneveld's heated length factor, $K_4$ (Eq. \ref{Eq:K4}). Instead of $K_4$, the Bobkov heated length factor, $k_3$ (Eq. \ref{Eq:HL}), has been used with the Groeneveld LUT (Figure \ref{Fig:GB-LUT}). Consequently, the performance of the Groeneveld LUT is greatly enhanced: The CHF decreases more smoothly along the channel, the NMAE is reduced to 0.25\%, and the Groeneveld LUT reproduces the Bobkov LUT.

\item Comparing Tables \ref{Tab:ValRanges} and \ref{ASD}, it can be observed that the operating pressure range of VVER-1000 (15.5$-$15.6 MPa) is outside the corresponding validity range of the Biasi correlation (0.27$-$14 MPa). Therefore, the Biasi correlation is not suitable for use with VVER-1000 rod bundles. It is only considered here for being a historical standard. The Biasi correlation underestimates the CHF by an NMAE of 15.03\%. Other research that reports on the Biasi correlation include:
\begin{itemize}
\itemsep -0.3 em
\item Kumamaru et al. \cite{Kumamaru1989} report that the Biasi correlation overestimates measured CHFs by factors of 10 or 100 at pressures from 3 to 12 MPa, mass fluxes from 20 to 410 kg/(m$^{2}$s), and inlet qualities from 0.4 to 0.9.
\item Yoder et al. \cite{Yoder1985} also report that the Biasi correlation overestimates the CHF for saturated qualities sometimes by an order of magnitude for pressures between 3.8 and 13.4 MPa.
\item Kao and Kazimi \cite{Kao1983} report that the Biasi correlation slightly overestimates the CHF in GE 9-rod bundles for critical qualities between 0.151 and 0.4885 at a pressure of 6.9 MPa.
\item When used computationally, the TRACE V5.0 Theory Manual \cite{TRACE} reports that for the whole range of saturated qualities, the Biasi correlation significantly overestimates the LUT CHFs at a pressure of 0.3 MPa, and mass flux of 100 kg/(m$^{2}$s), whereas it nearly reproduces the Groeneveld LUT CHF values at a pressure of 7 MPa, and mass flux of 1000 kg/(m$^{2}$s) over the same range of saturated qualities. Merging Kao and Kazimi's results \cite{Kao1983}, the TRACE results \cite{TRACE} and our results, it appears that as the pressure is increased, the Biasi correlation moves from overestimation (at 0.3 MPa) to a nearly perfect reproduction of CHF values (at 6.9$-$7 MPa) to underestimation (at 15.5$-$15.6 MPa).
\end{itemize}

\item Both the Bowring and W-3 correlations reasonably overestimate the CHF, which agrees with the results reported by Kao and Kazimi \cite{Kao1983}, Yoder et al. \cite{Yoder1985}, and Hejzlar and Todreas \cite{Hejzlar1996}. This consistency is due to the Bowring and W-3 correlations taking account of inlet conditions. The Bowring correlation shows a slightly better performance than the W-3 correlation because it incorporates a full heat balance. As stated previously, the MDNBR is, in general, a relative value that depends on the specific correlation used. A rough estimate of the MDNBR value specific for both Bowring and W-3 correlations can be given if we assumed that \textit{(a)} the boiling crisis occurs exactly at the channel outlet, and that \textit{(b)} at the channel outlet, the heat flux equals the CHF as calculated by the Bobkov LUT divided by 1.3 (i.e., $q = \mathrm{CHF}_{\mathrm{Bobkov}} / 1.3$). If this value of the heat flux is used with CHF values calculated by the Bowring and W-3 correlations, the MDNBR will be equal to 1.83 and 1.96 for the Bowring and W-3 correlations, respectively. In other words, an MDNBR = 1.3 calculated by Bobkov LUT is equivalent to an MDNBR = 1.83 calculated by the Bowring correlation, and an MDNBR = 1.96 calculated by the W-3 correlation. This estimate for the W-3 MDNBR agrees within 9.68\% with the MDNBR value calculated by Shuffler et al. \cite{Shuffler2009} (i.e., MDNBR = 2.17), who implemented the W-3 correlation in the VIPRE subchannel code for a typical PWR.

\item The OKB-Gidropress correlation overestimates the CHF by nearly a factor of 2 over the whole length of the channel and exhibits the poorest performance of all CHF predictors. Mozafari et al. \cite{Mozafari2013} have found that the OKB-Gidropress correlation predicts a CHF value higher than that predicted by the W-3 correlation which agrees with our result. However, reportedly based on one of the datasets on which the Bobkov LUT is based, the OKB-Gidropress correlation deviates significantly from the predictions of the Bobkov LUT. Considering the correlation can only be found in one reference \cite{Mozafari2013}, the authors suspect that this discrepancy might be due to a typographical error made while reporting the correlation.
\end{itemize}

\subsection{Interpolation algorithms}

\begin{figure}[h!]
\centering
\includegraphics[width=\textwidth]{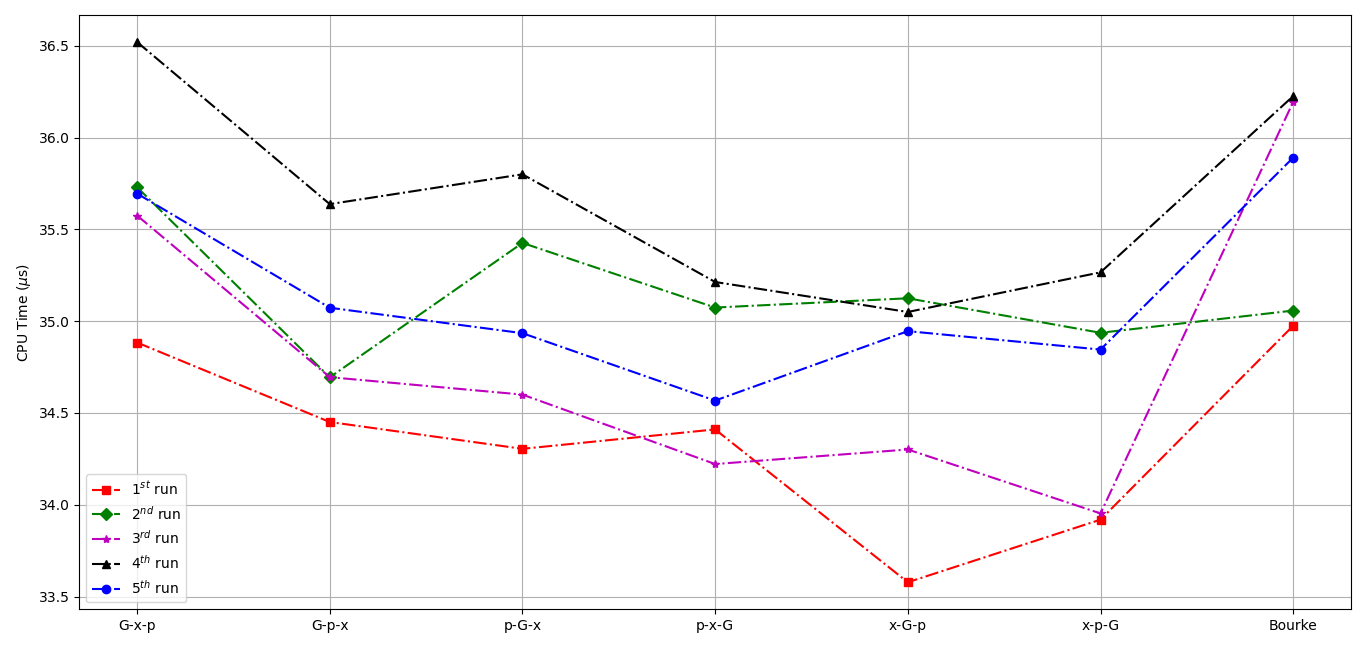}
\caption{Average CPU time for both the standard and Bourke trilinear interpolation algorithms}
\label{Fig:mean}
\end{figure}

\begin{figure}[h!]
    \centering
    \includegraphics[width=\textwidth]{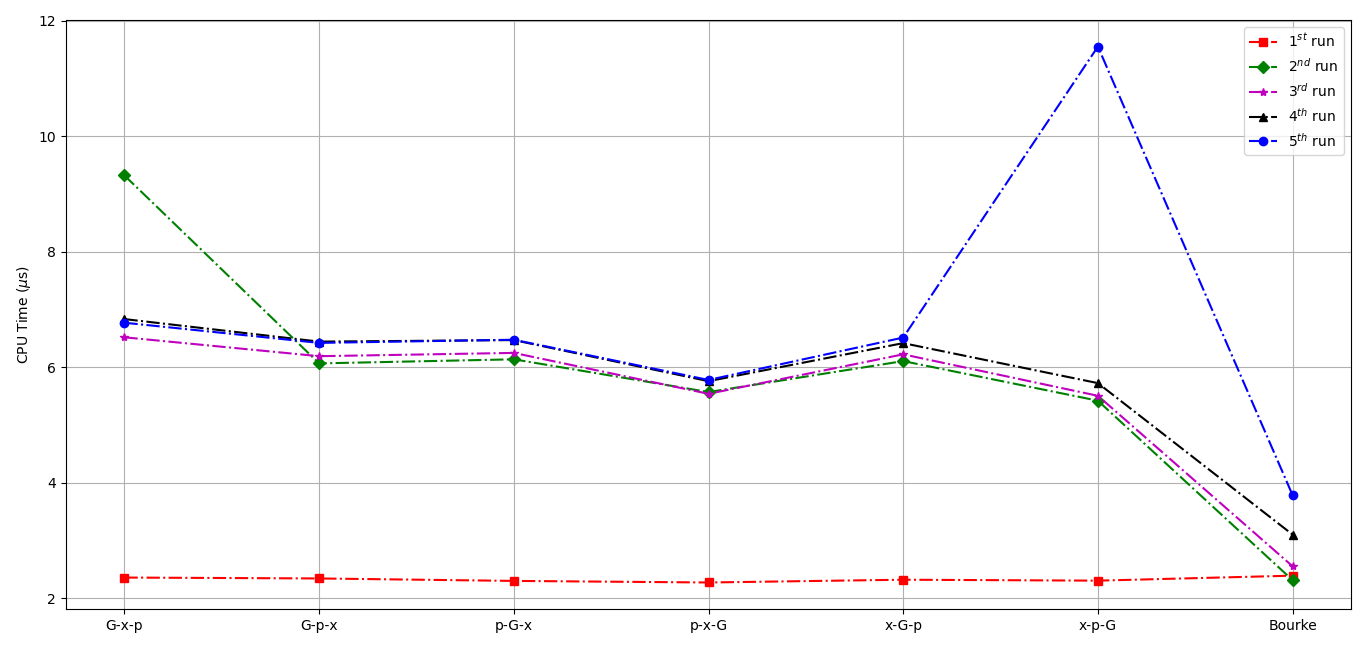}
    \caption{Standard deviation in CPU time for both the standard and Bourke trilinear interpolation algorithms}
    \label{Fig:STD}
\end{figure}

To assess time complexity, the algorithms were implemented in a Python code for 10$^6$ iterations and the Python code itself was run 5 times (i.e., 10$^6$ iterations per single run). The processor is an Intel Core i5-6200U CPU (2.3 GHz $\times$ 2) and the operating system is Linux Mint 20.1 "Ulyssa" Cinnamon Edition. The average CPU time and its standard deviation were calculated for the 5 runs, and the results are shown in Figures \ref{Fig:mean} and \ref{Fig:STD}, respectively. As observed in Figure \ref{Fig:mean}, all the algorithms nearly show the same average CPU time usage. While the Bourke algorithm shows a slightly larger average CPU time, its CPU time standard deviation is consistently smaller, which means it is less affected by noise \cite{Kube2013}. Because Python is known for its poor computational efficiency, only relative values of CPU times are relevant. In other words, the absolute values of average CPU times may be quite different would the algorithms be implemented in a compiled language like C++, and with different processors and operating systems. However, relative CPU times would exhibit the same general characteristics, and the Bourke algorithm would still be less affected by noise. While the RELAP-SCDAPSIM MOD4.0 code uses the $p$-$G$-$x$ arrangement of the standard algorithm \cite{RELAPModels}, we recommend implementing the Bourke algorithm in future versions.

\section{Conclusions and recommendations}

\begin{enumerate}
    \itemsep -0.3 em
    
    \item Axial heat flux non-uniformity does not affect the CHF in subcooled conditions. On the contrary, the heated length effect is responsible for the decaying trend of the CHF in the $q$-$z$ space and should be accounted for whether LUTs or empirical correlations are used to predict the CHF.
    
    \item The Groeneveld heated length factor ($K_4$ of Eq. \ref{Eq:K4}) is not suitable for subcooled conditions and exhibits a nearly abrupt decline at channel inlet. When estimating the CHF in VVER rod bundles, it is recommended to use the Bobkov heated length factor ($k_3$ of Eq. \ref{Eq:HL}) with both the Groeneveld and Bokbkov LUTs, as well as empirical correlations. Indeed, when the Bobkov heated length factor is applied to the Groeneveld LUT CHF values, it is found that the curves of the Groeneveld and Bobkov LUTs are nearly identical.
    
    \item The Groeneveld exponent $n=0.5$ will be used with the Groeneveld diameter factor ($K_1$ of Eq. \ref{Eq:K1}) when predicting the CHF in VVER rod bundles. Although the Tanase and Wong exponent correlations give better predictability over the whole range of parameters covered by the Groeneveld LUT in round tubes, the Groeneveld exponent appears to be more suited for VVER rod bundles.
    
    \item Both the Bowring and W-3 correlations consistently exhibit a reasonable overestimation of the CHF in VVER rod bundles. They can be implemented in safety codes to determine the MDNBR limit for rod bundles provided that the value of MDNBR at 112\% of nominal power is 1.83 for the Bowring correlation, and 1.96 for the W-3 correlation. Rough estimates of the MDNBR limit at other values of overpower (e.g., 118\% of nominal power) can be evaluated based on the methodology used in this paper. On the other hand, the Biasi and OKB-Gidropress correlations are not recommended for use with VVER rod bundles. The Biasi correlation is more suited for use at pressures around 7 MPa.
    
    \item The standard and Bourke trilinear interpolation algorithms nearly exhibit the same average CPU time. The Bourke algorithm, however, is less affected by computer noise.
    
    \item More research is needed to increase the number of data points in the Bobkov LUT, and the ``slice'' method should be applied to the table to ensure smoothness. It is also recommended to use the Bolshakov spacer grid factor ($F_B$ of Eq. \ref{Eq:FB}) with the Bobkov LUT instead of $k_4$ of Eq. \ref{ASD2}.
    
    \item The Groeneveld LUT nearly predicts the Bobkov LUT CHF curve when used with the Bobkov heated length factor. However, further research is needed on its correction factor formulas. Because correction factors depend on both geometrical and physical parameters, we recommend that a set of correction factors should be derived for each type of nuclear reactor. Deriving a set of correction factors that are expected to work with all types of nuclear reactors is, in our opinion, an elusive task. Additionally, the effect of unbalanced assemblies, expressed by $k_5$ of Eq. \ref{k5}, should be adopted in the implementation of LUTs in safety codes. This approach can enhance safety code prediction of boiling crisis at peripheral subchannels; a capability that has only been accessible to subchannel codes.

\end{enumerate}

\section{Acknowledgements}

The authors would like to show their utmost gratitude to Innovative Systems Software (ISS) for providing the VVER-1000 input deck without which the research conducted in this paper would not have been possible. The schematics in this paper have been generated by \textit{diagrams.net} to which we are also grateful.

\printbibliography

\end{document}